\documentclass{article}
\usepackage{arxiv}
\usepackage[utf8]{inputenc} 
\usepackage[T1]{fontenc}    
\usepackage{hyperref}       
\usepackage{url}            
\usepackage{booktabs}       
\usepackage{amsmath,amsfonts,amsthm}
\usepackage{nicefrac}       
\usepackage{microtype}      
\usepackage{cleveref}       
\usepackage{lipsum}         
\usepackage{graphicx}
\usepackage{natbib}
\usepackage{doi}

\usepackage{float}
\usepackage{graphicx}
\usepackage[english]{babel}
\usepackage{caption}
\usepackage{subcaption}
\usepackage{rotating}
\usepackage{pdflscape}
\usepackage{afterpage}
\usepackage{capt-of}
\usepackage{booktabs}
\usepackage{threeparttable}
\floatstyle{plaintop}
\RequirePackage{natbib}
\usepackage{threeparttable}
\usepackage{rotating}
\usepackage{hyperref}
\hypersetup{
  colorlinks=true,
  linkcolor=blue,
  citecolor=blue,
  urlcolor=magenta,
  hyperfootnotes=true,
}

\usepackage{enotez}
\let\footnote=\endnote
\usepackage{graphicx}
\usepackage[title]{appendix}
\usepackage{color}
\usepackage{graphicx,multicol}

\title{Cognitive and non-cognitive efficiency gaps between private and public schools in the Latin America region---a hybrid DEA and machine learning approach based on PISA 2022}

\date{}

\usepackage{authblk}

\newbox{\orcid}\sbox{\orcid}{\includegraphics[scale=0.06]{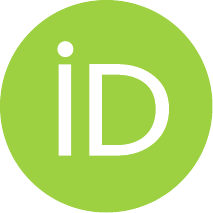}}
\author[1,2]{%
	\href{https://orcid.org/0000-0001-9333-8331}{\usebox{\orcid}\hspace{1mm}Marcos Delprato\thanks{Email: \texttt{md2645@bath.ac.uk} and \texttt{marcos.a.delprato@gmail.com}.}}%
}
\affil[1]{Department of Education,  University of Bath, UK}
\affil[2]{Instituto de Investigaciones Educativas, Universidad Nacional de Chilecito, Argentina}

\renewcommand{\shorttitle}{Efficiency gaps between private and public schools in the Latin America region}

\hypersetup{
pdftitle={Cognitive and non-cognitive efficiency gaps between private and public schools in the Latin America region---a hybrid DEA and machine learning approach based on PISA 2022},
pdfsubject={econ.GN},
pdfauthor={Marcos Delprato},
pdfkeywords={school efficiency, private and public schools, DEA, explainable machine learning, Latin America, PISA 2022},
}

\begin{document}
\maketitle

\begin{abstract}
Latin America's education systems are fragmented and segregated, with substantial differences by school type. The concept of school efficiency (the ability of school to produce the maximum level of outputs given available resources) is policy relevant due to scarcity of resources in the region. Knowing whether private and public schools are making an efficient use of resources --and which are the leading drivers of efficiency-- is critical, even more so after the learning crisis brought by the COVID-19 pandemic. In this paper, relying on data of 2,034 schools and nine Latin American countries from PISA 2022, I offer new evidence on school efficiency (both on cognitive and non-cognitive dimensions) using Data Envelopment Analysis (DEA) by school type and, then, I estimate efficiency leading determinants through interpretable machine learning methods (IML). This hybrid DEA-IML approach allows to accommodate the issue of big data (jointly assessing several determinants of school efficiency). I find a cognitive efficiency gap of nearly 0.10 favouring private schools and of 0.045 for non-cognitive outcomes, and with a lower heterogeneity in private than public schools. For cognitive efficiency, leading determinants for the chance of a private school of being highly efficient are higher stock of books and PCs at home, lack of engagement in paid work and school's high autonomy; whereas low-efficient public schools are shaped by poor school climate, large rates of repetition, truancy and intensity of paid work, few books at home and increasing barriers for homework during the pandemic.
\end{abstract}



\newpage
\section{Introduction}
\label{section1}

The learning crisis in the Latin American and the Caribbean (henceforth: LAC) region is worrying (\citealp{arias24a}), even more so when this regional learning crisis is compounded with wider inequality (\citealp{asadullah23,bracco25,fernandez24}) leading to an increasing school segregation (\citealp{gomez22}) in education systems which were already underfunded and resource-constrained (\citealp{dufrechou16}). In a study for the LAC region, \cite{arias23} estimate that 75\% of the students are low performers in math (not reaching the basic competencies given by level 2), and 55\% of students lacks basic reading skills (failing to understand a simple text). Within this adverse context, the COVID-19 pandemic, in turn, particularly aggravated the situation (\citealp{neidhofer21}) because the LAC region, with low levels of connectivity, experienced the longest period of school shutdown (270 days on average between 2020-2012; \citealp{bracco25}), and a sharp decline in learning (loosing around lost around 0.9-1.1 years of schooling; \citealp{azevedo21}). This education disruption affected relatively more vulnerable students (\citealp{brunori24}), generally attending public schools, who experienced the greatest losses. For instance, \cite{bracco25} carry out a simulation for the LAC region estimating a mean loss years of education for a student in the bottom decile of the income distribution of 81\%, nearly four times as large than a student located in the top decile (losing 22\%).

In this precarious reality of LAC education systems, school efficiency (i.e., achieving the maximum educational output for a given level of resources) is a very fitting policy concept to evaluate their long-term financial sustainability, particularly in a scenario of tight educational budgets and lower government educational investments. Efficiency on education provision, where schools make the best use of available inputs, had become a topic of acute debate among educational stakeholders (\citealp{dewitte17}), especially when there are few available resources which need to be efficiently administered. The LAC case, with the dual private-public nature of its education system (\citealp{elacqua18}), has shown a consistent learning divide in favour of students attending secondary private schools (\citealp{castro17,pagliarani25}). When it comes down to efficiency gaps, though, private schools' key elements (such as having more advantage students population and so lower costs of academic support without the need of remedial classes, and more resources and higher autonomy) may lead to a more proficient production process when transforming inputs into educational outputs than in public schools (\citealp{agasisti24}). Still, whether or not a differential in the production of educational outcomes by school type in the LAC region holds remains to be answered at a regional level, even more so after the pandemic. The only previous work in that direction is \cite{delprato21a}, but this study is based on pre-pandemic data (PISA for development, year: 2017) and relying on just four LAC countries.

This paper fills this gap by estimating private schools and public schools sub-equilibriums from an efficiency perspective using newer post-pandemic data which allows to capture, among other things, how standard schools' environmental features (like autonomy and selectivity and socioeconomic background of students) and new key drivers for remote learning suddenly brought about by the onset of the COVID-19 pandemic (such as the degree of digital preparedness; \citealp{xu25}) are empirically relevant as explanatory factors for differential on private and public schools levels of efficiency. Because the pandemic had severe impacts on psychosocial health and well-being of students (\citealp{caycho23,li25}), which were crucial for pedagogic continuity of students during remote learning (\citealp{delprato25b}), further to estimating the level and determinants of cognitive efficiency based on academic outputs, I also estimate private and public schools efficiency using non-cognitive outcomes.

Specifically, using data of nine LAC countries and a sample of 2,034 schools from PISA 2022, first, I estimate separately for private and public schools their cognitive and non-cognitive efficiency levels using Data Envelopment Analysis (DEA) (\citealp{johnes04,thanassoulis16}) and, second, I gauge the importance of determinants of school efficiency relying on recent interpretable machine learning methods (IML), i.e. the SHAP method (\citealp{lundberg17}) which allows to simultaneously assess a large number of efficiency drivers. As far I am aware, this is the first study which offers evidence at a regional scale on school type efficiency gaps for LAC and, methodologically, I expand the rather small and incipient literature of DEA plus IML (e.g., \citealp{boubaker25}) towards the school's efficiency domain. The paper attempts to answer the following research questions for the LAC region:

\begin{itemize}
  \item What is the degree of cognitive and non-cognitive efficiency by school type? If any, what is the extent of the school's efficiency gap by private and public schools?
  \item Which generic students and schools characteristics, as well as policy factors, are associated with the two kinds of efficiency by school type? And, what are the leading drivers of highly efficient and low efficient schools?
\end{itemize}

The remaining of the paper is structured as follows. Section \ref{section2} includes a review of school efficiency by school type and, in Section \ref{section3}, I introduce the data employed, the school efficiency model (DEA) specification, and summary statistics. The combined hybrid empirical approach I follow, i.e., DEA and interpretable machine learning to asses determinants of school efficiency, is described in Section \ref{section4}. Section \ref{section5} shows the results. Section \ref{section6} contains a discussion of results and concluding remarks.


\section{Literature review}
\label{section2}

Given the self-selection of wealthy students into well-resourced private schools as their families can afford the cost of high enrolment and, at the same time, high performing students from advantaged families being more prone to opt out from their assigned public schools by residential mobility (\citealp{delao23,elacqua18,pagliarani25}), which in turn increase wider school segregation (\citealp{bonal20,murillo17,gutierrez20}), it is not surprising that numerous studies based on international learning surveys have consistently shown that there is a persistent achievement gap across time between private schools and public schools students in the LAC region.

Based on PISA 2000 and 2012 waves for six LAC countries, \cite{delprato15}'s multilevel inequality analysis shows that, even after accounting by student's and school's factors and the heterogeneity of within and between-school SES inequalities, the math (reading) performance average gap for achieving basic knowledge$/$level 1 (or above) is in favour of private school students by around 26\%-60\% (66\%-96\%) than their counterparts. Likewise, based on PISA 2012 for 8 countries, \cite{castro17}, using a robust methodology (accounting for observables through the Oaxaca-Blinder decomposition and instrumental variables) find gaps between 64-67 points in maths, reading and science performance of private school students, and with cross-level covariates (individual, family, and school) accounting for around 85\% of this gap.

More recently, using PISA 2022 and the same sample of nine LAC countries as the current paper, \cite{pagliarani25}'s analysis of high performing public schools indicates how private schools over-performer public schools, i.e.,  around 83\% of students in the region attends public schools (the largest enrolment rate), though just 60\% of these reach the top tercile of the achievement distribution, whereas out of the 17\% enrolment of private school students, 41\% falls into the top tercile. Another recent study using PISA 2022 for the LAC region,  i.e., \cite{arias24a}, also finds a general positive private-public achievement gap trend in the region: ``accounting for students' and schools' socio-economic status, math scores remained higher in private schools than in public schools in seven LAC countries'' (p. 25).

Yet, are private schools more efficient --obtaining higher aggregated learning gains at a given level of resources-- than public schools due a more 'productive' linkage between educational inputs and outputs or, alternatively, is the superior performance of private school students due schools just being better-resourced? Previous school efficiency research is fairly mixed insofar as the evidence considered is from different regions' education systems. Country-wise in Europe, \citealp{mancebon12} find that higher levels of efficiency in public high schools than in subsidised private schools in Spain (using PISA data, year: 2006) once in students' backgrounds, school resources and individual management inefficiencies are removed. The same pattern against private Spanish schools is found by \cite{crespo14}; whereas \cite{salas20}'s study points in the reverse direction with private subsidised schools being more efficient than public schools.
For richer OECD countries, \cite{agasisti18}'s study for 30 countries based on PISA (2012) obtains a small negative gap given by the coefficient of the private school dummy in comparison to other determinants of efficiency; though, when the analysis is shifted towards developing countries from distinct regions (28 countries, PISA 2012), the same authors find the efficiency levels are higher in private schools compared to public schools (\citealp{agasisti19}). In the same vein, \citealp{delprato21b}, focusing on PISA for development (PISA-D) for seven low-middle and low income countries (i.e., Cambodia, Ecuador, Guatemala, Honduras, Paraguay, Senegal and Zambia), obtain a whole sample's positive gap on efficiency for private schools (which hold across most countries individually).

As far as the LAC's evidence as a region$/$unit of analysis is concerned, private-public efficiency studies are rather scarce as mostly evidence is at country level and at either primary level (e.g., \citealp{antequera19,vera08}) or for public schools (\citealp{agasisti22}). \cite{agasisti23}, using PISA 2018  data for 2,757 schools from 10 countries assess the impact of ICT availability and its use on school efficiency, find that out of the ten countries, private schools are more efficient than public schools --in Argentina, Brazil and Chile-- while the reverse association was found in Mexico and Peru (and having not impact on the remaining five countries). In a comparison exercise of efficiency levels and their drivers before and after the COVID-19 pandemic for LAC (using PISA 2018 and 2022 and 9 countries), \cite{delprato25} find that private and urban schools are more efficient than their counterparts, with the impact of private school effects on efficiency being larger for the 2022 wave in comparison to the 2018 wave (i.e.,0.0217 versus 0.0182) (also, see \citealp{delprato2025c}).

It should be stressed that both articles are not a regional study as such and school type production functions from the DEA analysis are not estimated separately but the private school effect is modelled in a second stage regression as a covariate (a private school dummy). Instead, \cite{delprato21a} do estimate separate production functions by carrying out an efficiency analysis for each school type using a PISA-D sample of four LAC countries: Ecuador, Guatemala, Honduras and Paraguay. In this paper, the authors obtain a positive efficiency gap for private schools of 6\% (private school efficiency of 0.88 and for public schools of 0.82), suggesting that the lower efficiency in public schools could be explained by the additional obstacles they face (e.g., higher prevalence of student work, higher rates of Indigenous students, smaller rates of preschool attendance).

Nevertheless, at the country level (and for secondary level) there is more evidence and, to some extent, results concerning the direction of school type efficiency are mixed.\footnote{For further details see, \cite{agasisti24}.} \cite{arbona22} who analyse the changes in productivity for around 4,600 Colombian schools using as output tests results at the end of secondary school (Saber 11) find that, on average, there is a better performance change in the public than the private sector, although there is a general deterioration in the education system. \cite{dejorge18} also for Colombia (using PISA 2012 and a meta-frontier approach), find that public schools are more inefficient than private schools, whilst the study of \cite{henriques22} relying on PISA-D for Ecuador, identify a larger proportion of public schools among top efficient schools (20 out 28) and small share of efficient private schools.

Taken the above review as a whole, although it points towards a positive private-public school efficiency gap (at least when the evidence is judged from a regional LAC perspective), further analyses should be carried out to establish whether this private-public efficiency divide still holds in a new context and time (namely: in the LAC region post-pandemic). In this new context, there has been a reshape of resources allocations shifting equilibriums of education sub-systems generated by the pandemic, with private schools being better prepared to support learning continuity (\citealp{arias24a}); students attending public schools being more likely to be less pedagogically engaged than those attending private schools (\citealp{delprato25b}); and private to public school migration in certain LAC countries (\citealp{elacqua25}). Established factors found in the literature driving efficiency gaps by school type (e.g., socioeconomic background of students, higher level of autonomy; \citealp{agasisti24}), need to be validated. The current paper is an attempt to fill this research gap.


\section{Data}
\label{section3}

The analysis is based on the latest wave of PISA  which is a learning survey from the OECD assessing the learning performance and skills in reading, mathematics and science of secondary students of around 15 years old of age to meet real-life challenges (\citealp{oecd23a,oecd23b}). The working sample contains nine LAC countries: Argentina, Brazil, Chile, Colombia, Dominican Rep., Mexico, Panama, Peru and Uruguay.\footnote{I exclude two additional LAC countries from the analysis, i.e., Costa Rica (because its dataset does not contain the key covariate household SES used as input for the DEA formulation) and Paraguay (with missing outputs, i.e., non-cognitive outcomes).} Since the analysis is at school level (measurement of efficiency), I construct averages of key variables from the students' and schools' dataset and, after merging them, the resulting working sample includes 2,034 schools, out of which 1,548 are public schools and 486 are private schools.

Importantly, and due to the sudden onslaught of the COVID-19 global pandemic where PISA had to be postponed for one year (that is, to 2022), additional questions were added to contextual questionnaires related to the pandemic and its new model of delivery of education through remote learning (\citealp{oecd23b}). This new component allows to gather information on school closures' association with performance (\citealp{jakubowski25}), the impact of the pandemic on students' wellbeing, belonging and engagement (\citealp{li25}), the roles of digitalisation in educational systems for remote learning (\citealp{xu25}) and support policies during the pandemic (\citealp{crato25}). All in all, this additional information is vital to link estimated levels of school efficiency in the LAC region by school type with either enablers or barriers for remote learning of students as explanatory factors.

\subsection{DEA specification}
\label{section31}

Given the depth of the information of PISA, including on outputs and inputs for efficiency studies (such as learning scores and household wealth and school infrastructure) and environmental factors, there are various DEA-school efficiency studies relying on PISA data from earlier waves of PISA (e.g., \citealp{agasisti23,agasisti18,aparicio22,mujiya25}) and, recently based PISA 2022, there are just a few studies for European countries (for example: \citealp{cordero25}) . Yet, there is a lack of DEA studies for the LAC region looking at school type efficiency gaps based on post-pandemic data. The DEA specification I follow mimics an education production where inputs (I) are converted into outputs (O) under a given technology (\citealp{hanushek20}). The selection of I and O, in turn, is guided by previous studies on school efficiency (for a review, see: \citealp{dewitte17}).

Specifically, the analysis uses two production functions, one for academic outputs (school average values for students’ scores on math, reading and science) and the other formulation uses a non-academic output.\footnote{The continuous non-cognitive outcome is compositive indicator obtained through factor analysis measuring overall well-being based on seven domains$/$personality traits (i.e., assertiveness, cooperation, curiosity, emotional control, empathy, perseverance, and stress management), with each domain having different questions. The construction of this outcome condensing (positive) personality traits are important due to its positive impact on academic achievement (\citealp{almlund11}) as they are found to be protective factors against adversity (\citealp{windle11}).} Further, as inputs I employ four indicators capturing distinct dimensions of school resources, i.e., school (physical and educational) infrastructure, the student-teacher ratio (STR), and the school average of household wealth of students. Summary statistics for outputs and inputs are shown in Table \ref{table1}, underscoring the segregated nature of school systems in the LAC region. There is a positive private-public schools gap both for outcomes and inputs (Table \ref{table1}, column 5), with private schools having higher average performance on academic outputs scores (in the range of 59-70 points) and non-academic outputs (of 0.07), and statistically significant differences for inputs on family wealth (so contextual poverty in private schools is lower) and schools' infrastructure.

\medskip
\begin{center}
  [Table \ref{table1} here]
\end{center}
\medskip

\subsection{School efficiency determinants}
\label{section32}

Table \ref{table2} shows the array environmental factors employed to assess efficiency levels by school type (where nearly all of them are statistical different across the two sub-samples). Among students' determinants on efficiency, Panel A (Table \ref{table2}) shows that private schools are better off in terms of their observed students characteristics compared to public schools, which is in line with previous studies for the LAC region based on PISA and PISA-D (\citealp{castro17,delprato21a,pagliarani25}). Private school students have large number of books at home (= 0.78) with the likelihood of students working being lower (unpaid and paid work), private schools have repetition rates at primary or lower secondary considerable lower (by 10\%-11\%) and their students having a larger ICT stock at home and a more intense use of it, as well as better school climate and 27\% higher rate of using a digital device (DD) during the pandemic.

\medskip
\begin{center}
  [Table \ref{table2} here]
\end{center}
\medskip

Similarly, the superiority of private schools contextual features contrasts with those from public schools (Panel B, Table\ref{table2}). Private schools have fewer barriers to educational instruction (all indices for barriers related to materials, infrastructure and ICT are lower); also, private schools face wider competition for their places (gap = 0.22) and have more autonomy (gap = 1.39) than public schools, as well as more methods to assess teaching (gap of 0.20) and better social$/$well-being policies (gap of 0.10). As it is widely known in the literature (\citealp{delao23,elacqua18}), private schools operate through selection (particularly based on previous academic records of students, with a gap with respect to public schools of  20\%) which allow them to capture more affluent students (public schools have a 26\% larger rate of disadvantaged students). This dominance of schools' features and resources of private schools is also translated into a better preparation for online learning during pandemic, under a shorter period of closure (39 days shorter in private schools). As it can be seen in Panel C (Table \ref{table2}, column 5), for example, policy-wise private schools have increasing investments for teachers and students for remote learning (be it resources, times to adapt to new learning platforms, etc.), coupled with a higher rate of students attending distance learning.


\section{Methodological approach}
\label{section4}

\subsection{Estimation of school type efficiency--DEA}
\label{section41}

I rely on the nonparametric methodology Data Envelopment Analysis (DEA) (\citealp{johnes04,thanassoulis16}) to estimate school efficiency scores. Each school (or decision making unit, DMU) transforms inputs into outputs through a production process yielding a measure of technical efficiency (\citealp{farrell57}), where technical efficiency is an indicator computed in relationship to the estimated frontier, which represents the maximum output that each school can achieve given the available resources (\citealp{worthington01}). Since I carry out a separate analysis by school type, efficiency is measured with respect to the estimated frontier of each school type under different technologies (private and public schools), and for the two types of outputs (cognitive and non-cognitive outcomes); hence, four DEA models are estimated.

Formally,\footnote{Mathematical notation and concepts are adapted following \cite{badunenko16}, \cite{bolos23} and \cite{simar13}.} a school (or DMU) $i$ ($i$ = 1,...,$n$) consumes $m$ inputs and produces $s$ outputs through an associated educational activity (\textbf{x}, \textbf{y}), a pair of nonnegative vectors, with  inputs vector \textbf{x} = ($x_{1},..., x_{m}$) and the outputs vector ($y_{1},..., x_{s}$), and activities defined in  $\mathbb{R}_{>0}^{m+s}$. Let the set of all $n$ schools be $\mathcal{D} = \{\text{school}_{1},...,\text{school}_{n}\}$, and the production possible set $\mathcal{P}$ (or technology $\mathcal{T}$) is defined within $\mathcal{D}$ by educational activities given by:

\begin{equation}\label{PPS}
  \mathcal{P} = \big\{ (\textbf{x}, \textbf{y}) \in \mathbb{R}_{>0}^{m+s} \text{  } | \text{   } \textbf{x} \ge X\lambda, \text{   }  \textbf{y} \le Y\lambda, \text{   } \lambda \in \mathbb{R}_{+}^{n}  \big\}
\end{equation}

Considering two activities, it is said that one outperforms the other if it consumes less inputs while producing more outputs; that is, (\textbf{x}, \textbf{y}) is strictly dominated by (\textbf{x}', \textbf{y}') if \textbf{x}' $<$ \textbf{x} and \textbf{y}' $>$ \textbf{y}. Here one activity is equivalent to a one school. A school is efficient if it is not strictly dominated by any positive combination of schools in $\mathcal{D}$. The set of efficient activities (or efficiency schools) in $\mathcal{P}$ is known as the efficient frontier ($\mathcal{P}^{\partial}$), a subset of efficient points (schools) in $\mathcal{P}$ in the upper boundary of $\mathcal{P}$. The degree of school inefficiency is the distance from (\textbf{x}, \textbf{y}) to it, with technically inefficient schools operating at points in the interior of $\mathcal{P}$ and technically efficient schools operating somewhere along the technology defined by $\mathcal{P}^{\partial}$. Given the  set of reference schools $\mathcal{D}$, an efficiency score is function: $\theta = f$(\textbf{x}, \textbf{y}). I invert school's estimated efficiency scores which are  bounded by (0,1] so the further from one a score is, the greater the lack of efficiency of a school. For the four models, I tested constant returns to scale (CRS) against variable return to scale (VRS), leading --in all four cases-- to a rejection of the null hypothesis that the global technology is CRS.\footnote{Tests results are available from the author upon request.}

The output-based measure of technical efficiency for each school (i.e., $\theta_{i}$) is calculated by solving the following linear programming problem:

\begin{equation}\label{lineaprog}
\begin{split}
  \widehat{F}^{o}_{i}\big(y_{i}, x_{i}, y,x|\text{VRS}\big) & = \underset{\theta, z}{\text{max }} \theta \\
  \text{s.t.} & \sum_{i=1}^{n} z_{i} y_{is} \geq y_{is} \theta_{s}, \text{   }  s = 1,..., S \\
  & \sum_{i=1}^{n} z_{i} x_{im} \leq x_{im} \theta_{m}, \text{   }  n = 1,..., M \\
  \text{and  } &  z_{i} \geq 0, \text{   }  \sum_{i=1}^{n} z_{k} =1
  \end{split}
\end{equation}

where $z_{i}$ represents the process operating levels and the additional restriction $\sum_{i=1}^{n} z_{i} =$1 is added for Variable Returns to Scale (VRS) which were accepted against CRS. The $\widehat{\theta}_{i}$ estimates for the scores are obtained by solving Eq. \eqref{lineaprog} for each school type (and two set of outputs), obtaining $\widehat{\theta}^{\text{private, y}}_{i}$ and $\widehat{\theta}^{\text{public, y}}_{i}$ (for cognitive outputs), in addition to  $\widehat{\theta}^{\text{private, ync}}_{i}$ and $\widehat{\theta}^{\text{public, ync}}_{i}$ (for non-cognitive outputs).

\subsubsection{Stochastic dominance tests on school type's efficiency profiles}
\label{section411}

Once the set of efficiency scores for the four models are estimated, I test for stochastic dominance (SD) of efficiency scores by school type.\footnote{I use the \texttt{PySDTest} Python package (\citealp{lee24}).} That is, using the estimated (cumulative density) frontier for private schools and public schools I check if one is superior, in other words, to assess whether globally one school type is more or less efficient than the other type in the use of resources for a given level of outputs (first order SD), and second order SD adds the volatility concept to these cumulative densities comparison assuming that the educational planner is `risk-adverse' to it preferring lower heterogeneity across schools efficiency levels while minting the same level of efficiency. Specifically, I run first and second order dominance tests by assessing the following null hypotheses:

\begin{equation}\label{SDtest}
\begin{split}
  H^{s}_{0,1}  & = \widehat{\theta}^{\text{private, y}} \preceq_{s} \widehat{\theta}^{\text{public, y}} \\
  H^{s}_{0,2}  & = \widehat{\theta}^{\text{private, ync}} \preceq_{s} \widehat{\theta}^{\text{public, ync}} \\
  \end{split}
\end{equation}

for $s$ = 1 (first order SD) and $s$ = 2 (second order SD), and the alternative hypotheses $H^{s}_{1,i}$ (for $i$ = 1, 2) are the complement of the null hypotheses.

\subsection{Estimation of determinants of efficiency--interpretable machine learning}
\label{section42}

Once efficiency levels are estimated in the first stage (as shown in Section \ref{section41}), DEA studies proceed to a second stage regression using conventional econometric analytic models where environmental factors are used to predict the efficiency scores of DMUs, through a Tobit model or by using a bootstrap approach that yields valid inference (\citealp{daraio10,simar07}). Though, ML techniques and related interpretable machine learning (IML) approaches such as Shapley additive explanations (SHAP) value techniques (\citealp{lundberg17}), with better predictive power and robustness against multicollinearity issues and outliers, can improve the predictive results of two-stage DEA (\citealp{boubaker25,chen21}). The current study approach falls within emerging body of research combining DEA and ML techniques. Taking into account that the number of determinants of school efficiency assumed is large, i.e., 42 covariates (Table \ref{table2}), combining DEA with ML allows to better address and circumvent the curse of dimensionality of DEA with big wide data.

Rather than modelling (continuous) efficiency scores, I define a binary dependent variable for cognitive and non-cognitive efficiency ($\theta^{\text{binary}}_{i}$), taking the value of 1 if the estimated $\widehat{\theta}_{i}$ is above the regional mean, and 0 otherwise. Thus, the second stage ML analysis shows the predictive power of students' and schools' determinants for the probability that a school falls into the top half of efficient schools, either for the private or public school sample. This modelling approach is related to the idea that is more valuable finding out which factors pull a school’s efficiency towards the top of the efficiency frontier per school type, instead of finding out the factors behind the average school efficiency of the sample.

\subsubsection{Machine learning--models comparison}
\label{section421}

Since the school efficiency dependent variable is binary, three ML binary-classification models are used: logistic regression (logit), neural networks (NN) and gradient boosted trees (GBT) (\citealp{hastie09,kumari17}). The optimisation of hyperparameters uses a grid search to arrive to the best performing version of each model. Training of each model follows a stratified 5-fold cross validation with a 80\%-20\% random division of the dataset into training and testing sets, correspondingly. Following that, I select the final ML model relying on two performing metrics: the area under the receiver operator characteristic curve (AUROC), and the area under the precision recall curve (AUPRC) (\citealp{naidu23}). Moreover, each model has different specifications (6 for logit, 9 for NN and several for GBT).\footnote{In particular, for the logit model six specifications are considered using on two penalties terms (L1-Lasso regularisation and L2-ridge) and three inverse of regularisation strengths (C = 0.1, 1, 10); for the NN model, nine specifications are employed with three hidden layer size sets: \{200\}, \{100, 100\}, \{200, 100, 50\}, and three activation rules (relu, tanh, logistic); for the GBT model, I assess numerous specifications combining the number of estimators (100, 500, 1000, 5000), the sub-sample ratios (0.5, 0.7, 0.9), the trees of maximum depth (3, 5, 7, 9) and the three learning rates (0.001, 0.01, 0.1).} A comparison of the performance of the three ML models is shown in the Appendix (Table \ref{tableA1}). Estimated metrics show that, across the four ML models, there is a superiority of the GBT model in terms of performance. Taking the cognitive outcomes model for private schools as an example, the GBT's  AUROC and AUPRC scores are 0.769 and 0.855, whereas for logit these are 0.727-0.792 and for NN model 0.678 and 0.721. The same holds for remaining three models. Thus, I include below a description of the GBT model which will be used for initial estimations of school efficiency's determinants and these estimated values are then used to obtain the IML results based on the Shapley Additive Explanations (SHAP) method (\citealp{lundberg17}).

\subsubsection{Chosen method: gradient boosted trees}
\label{section422}

Gradient boosted trees\footnote{For the analysis, I use the XGBoost implementation of the GBT model (\citealp{chen16}).} is a ML method composed of iteratively trained decision trees and it is very efficient and flexible as the final ensemble of trees can capture non-linear and interaction effects between predictors (\citealp{krauss17}). A tree ensemble model uses $K$ additive functions to predict the output (\citealp{chen16}):

\begin{equation}\label{xgboost1}
  \widehat{y}_{i} = \phi(\mathbf{x}_{i}) = \sum_{k=1}^{K} f_{k}(\mathbf{x}_{i})
\end{equation}

Here, $\widehat{y}_{i}$ denotes the prediction label for a given sample $f(\mathbf{x}_{i})$, and $f_{k}(\mathbf{x}_{i})$ represents the predicted score for the sample, an independent tree structure $q$ and leaf weights $\omega$, and $f(\mathbf{x})$ denotes the value of a leaf (using $\omega_{i}$ to represent the score of the $i$-th leaf), and where $f_{k} \in \mathcal{F}$ is the space of regression trees: $\mathcal{F} = \{f_{k}(\mathbf{x}) = \omega_{q(x)}\} (q: \mathbb{R}^{m} \mapsto T,  \omega \in \mathbb{R}^{T})$, where $T$ is the number of leaves in the tree. To learn the set of functions used in the model, the following regularized objective is minimised:

\begin{equation}\label{xgboost2}
\begin{aligned}
  \mathcal{L}(\phi) &= \sum_{i} l( \hat{y}_{i}, {y}_{i}) + \sum_{k} \Omega (f_{k})    \\
  \text{and  } \Omega(f) &= \gamma T + \frac{1}{2} \lambda \|\omega\|^{2}
\end{aligned}
\end{equation}

where $l$ measures the distance between the prediction and the target dependent variable, and $\Omega(.)$ is the penalisation term minimising the risk of overfitting. An additive form for optimisation is used when training the model:

\begin{equation}\label{xgboost3}
  \mathcal{L}^{(t)} = \sum_{i=1}^{N}l\big(y_{i}, \hat{y}^{t-1}_{i}, {y}_{i}) + f_{t}(\mathbf{x}_{i})\big) + \Omega (f_{t})
\end{equation}

where $\hat{y}^{t-1}_{i}$ is the prediction of the $i$-th instance at the $t$-th iteration and $f_{t}$ is added to improve the model according to Eq. (\ref{xgboost2}).

\subsubsection{Explainable ML estimates: the SHAP method}
\label{section423}

Shapley values as a ML explanatory framework (\citealp{lundberg17}) is one of the most common approach within the IML field which has become more frequently used in educational research (\citealp{ersozlu24}). Shapley values have their origin in game theory (\citealp{shapley53}), measuring the average marginal contribution of a player in a cooperative game. In the context of ML, the player is interpreted as a characteristic or covariate (an attribute), and the cooperative game becomes the prediction task performed by the model. I follow the SHAP method (\citealp{chen23,lundberg17}) as an explanatory method of estimates of the GBT model. I estimate SHAP values' main effects for the whole range of covariates of Table \ref{table2}. SHAP values\footnote{I use the \texttt{TreeExplainer} package (\citealp{lundberg20}) which offers an exact calculation of SHAP values for tree-based models.} offer a local explanation of a model output by calculating the degree to which each feature (or variable) contributes to a given prediction value when conditioning on that feature. Formally, the change on prediction or SHAP value $\phi_{i}$(.) for feature (or covariate) $i$ in model $f$ for data point $x$ is given by:

\begin{equation}\label{shap1}
  \phi_{i}(f, x) = \sum_{R \in \mathcal{R}} \frac{1}{M!}\big(f_{x}(P^{R}_{i} \cup i) - f_{x}(P^{R}_{i})\big)
\end{equation}

where $\mathcal{R}$ is the set of all feature permutations, $P^{R}_{i}$ is the set of all features before $i$ in the ordering $R$, $M$ is the number of input
features, and $f_{x}$ is an estimate of the conditional expectation of the model’s prediction. The additive property of SHAP values means that SHAP values add to the output of the model:

\begin{equation}\label{shap2}
  f(x) = \phi_{0}(f) + \sum_{i=1}^{M}\phi_{i}(f, x)
\end{equation}

Attributions based on the Shapley interaction index result in a matrix of feature attributions. SHAP values main effects (on the matrix's diagonal) are obtained as the differences of SHAP values and the off-diagonal elements (i.e., SHAP interactions) for a given feature, and are defined as:

\begin{equation}\label{shap3}
  \Phi_{i,i} = \phi_{i}(f,x)-\underbrace{\sum_{j \neq i}^{}\Phi_{i,j}(f,x) }_{\text{SHAP interactions}}
\end{equation}


\section{Results}
\label{section5}

\subsection{DEA's estimates for private and public school efficiency}
\label{section51}

\begin{figure}[ht!]
\centering
\subfloat[\footnotesize{Histogram and density: cognitive outcomes}]{\includegraphics[width=.49\textwidth]{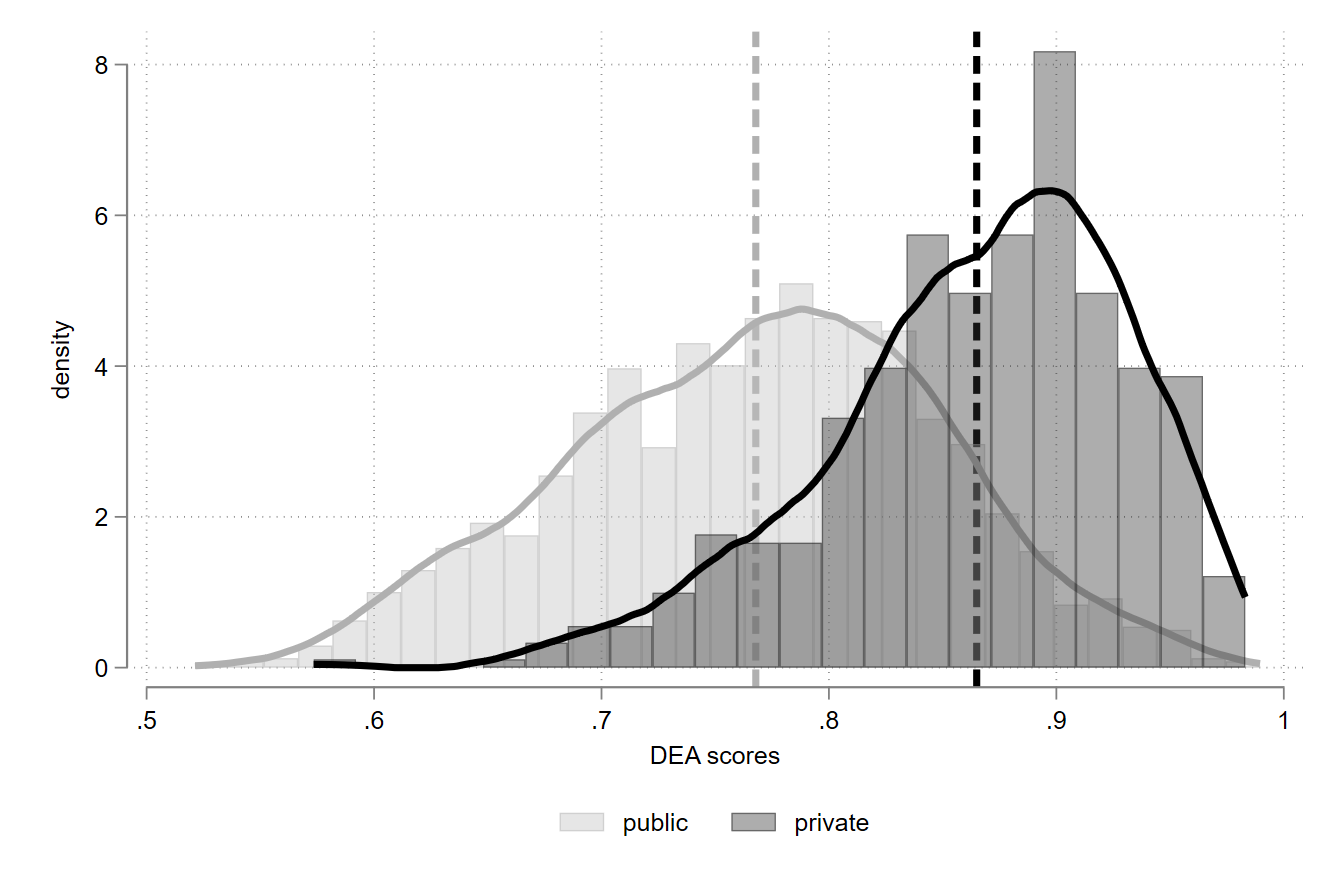}\label{figure1a}}
\subfloat[\footnotesize{Histogram and density: non-cognitive outcomes}]{\includegraphics[width=.49\textwidth]{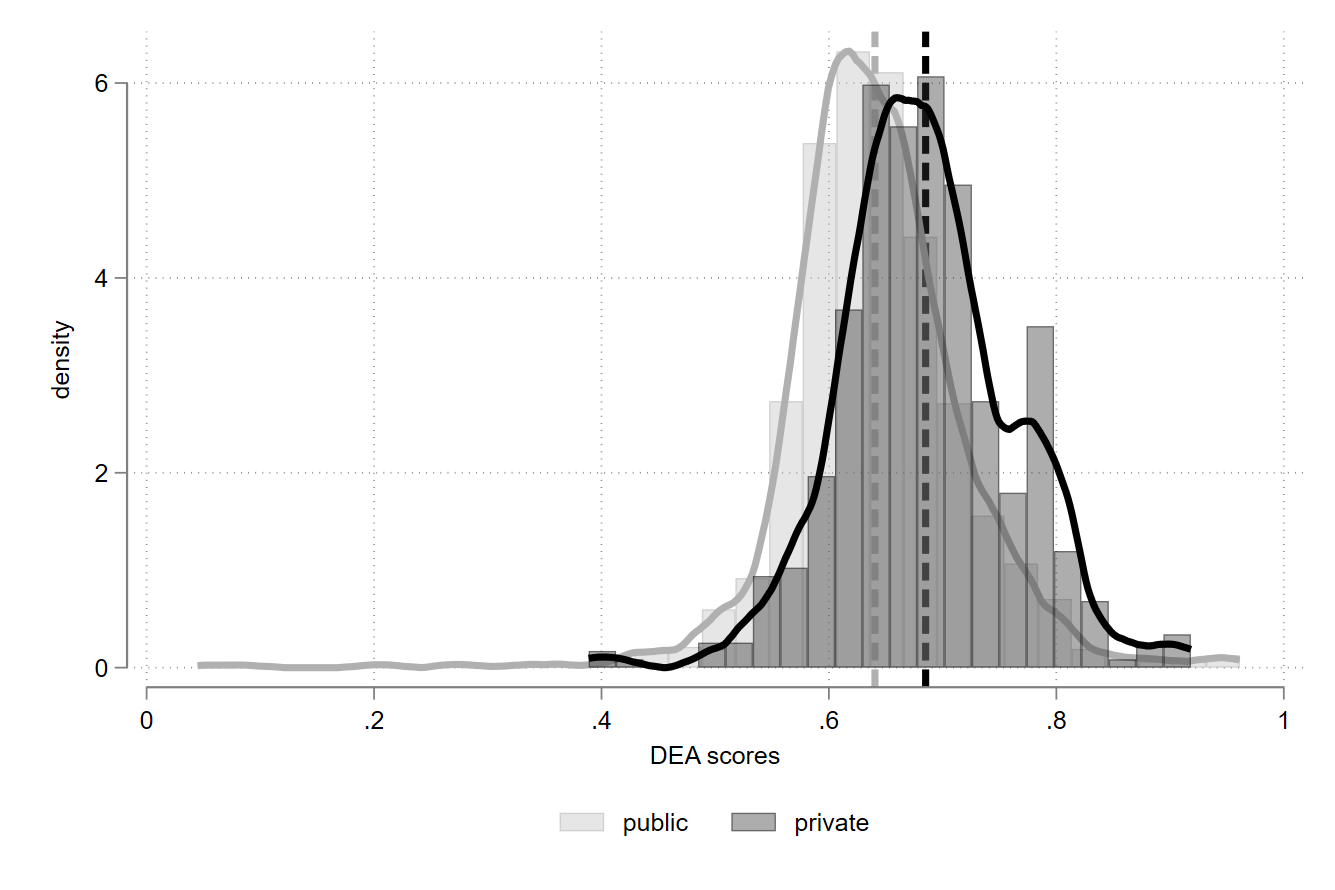}\label{figure1b}}
\\
\subfloat[\footnotesize{Countries' efficiency gaps}]{\includegraphics[width=.49\textwidth]{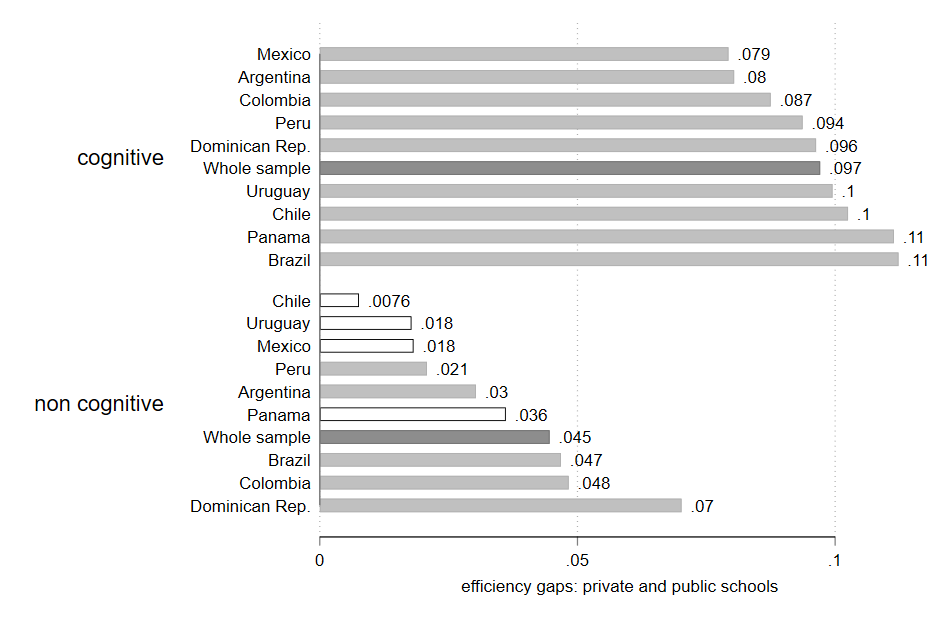}\label{figure1c}}
\subfloat[\footnotesize{Scatter efficiency cognitive and non-cognitive}]{\includegraphics[width=.49\textwidth]{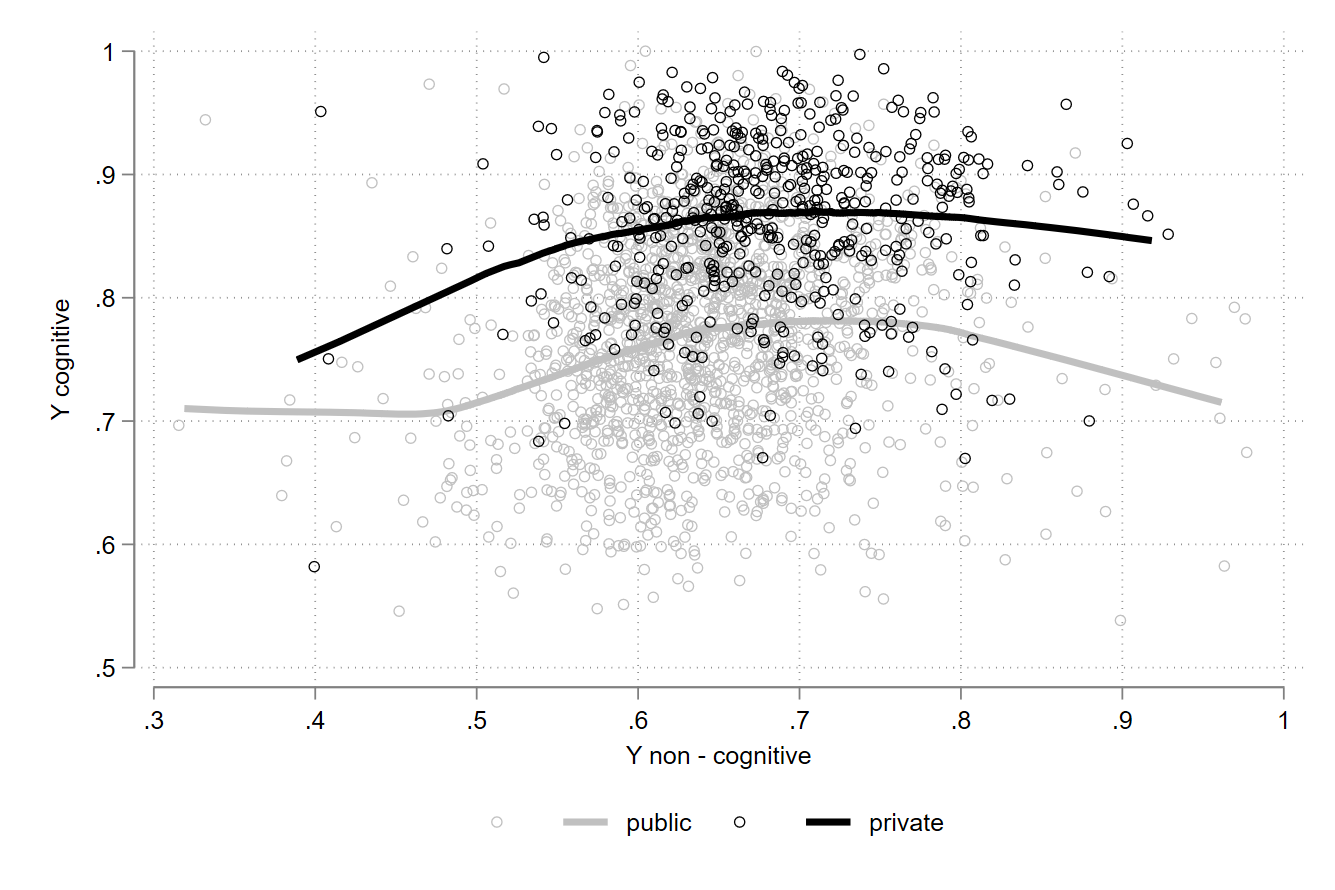}\label{figure1d}}
\caption{DEA estimated efficiency scores by school type and scatter of cognitive and non-cognitive outcomes}
\label{figure1}
\medskip
\begin{minipage}{0.99\textwidth}
\footnotesize{
Notes: (1) Country efficiency gaps bars in white are not statistically significant at 10\%.
}
\end{minipage}
\end{figure}

Estimated school efficiency scores obtained through the DEA analysis of cognitive outcomes by school type are displayed in Table \ref{table3}. Panel A contains estimates for the entire sample of private and public schools of technical efficiency (TE) and its bias-corrected bootstrap version with 95\% CI (TEBC) and bootstrap performance. The focus is on TEBC estimates (Panel A, column 2 of Table \ref{table2}). The mean efficiency score for private schools in the LAC region is 0.865, and for public schools, efficiency is much lower at 0.768 --a relative efficiency gap in favour of private schools of around 10 points. Hence, there is more room to increase efficiency (boosting average academic scores while keeping the stock of inputs constant) in public schools (by 23\%); and, in private schools, by 14\%. Lower efficiency goes in parallel to a wider dispersion in the case of public schools because the IQR of $\widehat{\theta}^{\text{public, yc}}$  is larger (= 0.117) than the for $\widehat{\theta}^{\text{private, yc}}$ (= 0.083) (Panels B and C). This wider dispersion on efficiency in public schools can also be clearly seen in Figure \ref{figure1a}, whilst the density of $\widehat{\theta}^{\text{private, yc}}$ is much narrower. As a result, cognitive outcomes school efficiency estimates indicate that public schools are not making the most of available resources given their lower technical efficiency in comparison to private schools.\footnote{A robustness DEA analysis, accounting for the presence of school outliers, is included in the Appendix (Figure \ref{figureC1} and Table \ref{tableC1}).}

\medskip
\begin{center}
  [Table \ref{table3} and Figure \ref{figure1} here]
\end{center}
\medskip

An inspection of countries' estimates (Panels B and C of Table  \ref{table3}) indicate a great gap between the most and the least efficient country (i.e., Uruguay against Dominican Rep.) of around 0.13 in terms of technical efficiency, both for private and public schools, highlighting specific country challenges. At the bottom bracket of school's efficiency$/$performance one can find Argentina, Panama and Dominican Rep. whose education systems, either public or private ones, are the worst performing, whilst Uruguay's private and public education systems are the most technically efficient. Among the remaining countries, however, efficiency ranks depends on school type: Brazil's and Chile's private school efficiency moves up into the ranking compared to rank of their public schools (from sixth to second and fourth to third, respectively), whereas for Peru, Colombia and Mexico the opposite holds. It is important to emphasise that the whole sample gaps on private-public school efficiency of 10 points are all positive and uniformly distributed across all countries, ranging between 0.08 and 0.11, with all these gaps being statistically significant (see: Figure \ref{figure1c}). Additionally, a closer look into countries' efficiency scores distributions,\footnote{In Appendix B (Figure \ref{figureB1a}) I include the density functions of DEA scores for private and public schools for each country. Besides their differential in these densities mean within each country, densities of efficiency scores have similar shape (and in, consequence, similar variation across the most efficient and less efficient schools within each school type) in the case of Colombia, Dominican Rep., Mexico, Panama and Peru. But, for the remaining countries in the LAC sample (i.e., Argentina, Brazil, Chile, and Uruguay), heterogeneity is much larger in the public school sectors, adding to the lower overall performance on technical efficiency, the complexity of having more inequality among the less and most efficient public schools relatively to private schools.} suggests that efficiency scores are much more dispersed within countries than between them, so that a comparison based on country averages would be misleading (Appendix B, Figure \ref{figureB1a}). Larger within-country disparity in public schools is linked to wider private-public efficiency gaps  (e.g., Brazil, Chile and Uruguay).

Table \ref{table4} contains DEA's results using as output the non-cognitive outcome (measuring overall well-being based on seven domains/personality traits). With estimated efficiency of $\widehat{\theta}^{\text{private, ync}}$ = 0.685, and $\widehat{\theta}^{\text{public, ync}}$ = 0.640 (Table \ref{table4}, column 2) for the whole LAC sample, the cognitive private-public efficiency gaps are translated to non-academic outputs as well, although with a smaller magnitude (of 0.045 rather than 0.10). Therefore, private schools are more technically efficient than public schools in two fronts: academic learning and socioemotional skills or soft skills. The lower levels of estimated technical efficiency based on the non-cognitive outcome compared to the one based on cognitive outcomes indicates that the average distance to the non-cognitive efficiency frontier is higher than for the education production function based on academic outputs, thereby showing a lower linkage of inputs with non-academic outputs. Put it differently, there is scope to increase non-cognitive scores and so efficiency by around 36\% (in public schools) and 31.5\% (in private schools) keeping the stock of inputs constant. These lower technical efficiency goes in tandem with a similar (and lower) between-schools heterogeneity for the distribution of efficiency scores by school type (IQRs of 0.085 and 0.090; Panel B of Table \ref{table4}). These lower schools' dispersion for soft skills estimated efficiency is clear when comparing the densities in Figure \ref{figure1b} (for $\widehat{\theta}^{\text{ync}}$) against Figure \ref{figure1a} (for $\widehat{\theta}^{\text{y}}$), with the latter being much narrower.

\medskip
\begin{center}
  [Table \ref{table4} here]
\end{center}
\medskip

When considering estimates at the country level (Panel B, Table \ref{table4}), in all nine countries soft-skills school efficiency is superior in private schools. Yet, because in four out of the nine countries private-public efficiency gaps lack statistical significance, this may hint that the role of school type is not as relevant in the transformation of inputs into non-cognitive outcomes as for cognitive outcomes. Specifically, with smaller private-public gaps than the whole sample (= 0.045), Chile, Uruguay, Mexico and Panama public schools are equally as efficient than private school in the production of soft skills, at least in statistical terms (see: Figure \ref{figure1c}).\footnote{An examination of the countries' densities of $\widehat{\theta}^{\text{ync}}$ shows equivalent shapes of efficiency distributions by school type (Appendix B, Figure \ref{figureB1b}); the only difference here with respect to $\widehat{\theta}^{\text{ync}}$ is that there are quite a few public schools whose estimated scores are well below the 0.5 threshold.}  Non-cognitive ranking efficiency for the nine countries remains constant only for the most efficient country in either school setting (i.e., Chile), but in countries such as Colombia and Brazil for example, they switch from third and fourth in the private schools sample to the last two positions in public schools sample (Panel B, Table \ref{table4}).

It is also worthwhile to examine how the two kinds of technical efficiency are related per school. In Figure \ref{figure1d} I plot for each school the cognitive efficiency score (y-axis) and the non-cognitive efficiency scores (x-axis). This figure shows a positive correlation between cognitive and non-cognitive efficiency, particularly in the range of $\widehat{\theta} \in$ [0.5, 0.75]. This correlation is somewhat larger in public schools ($\widehat{\rho}$ = 0.1501, p-value = 0.0000) than in private schools ($\widehat{\rho}$ = 0.0947, p-value = 0.0369), perhaps pointing towards a stronger role of personality traits of public school's students for achievement than their counterparts because of the contextual disadvantages of public schools.

\subsubsection{Stochastic dominance tests results}
\label{section511}

\begin{figure}[ht!]
\centering
\subfloat[\footnotesize{First order SD-cognitive outcomes}]{\includegraphics[width=.49\textwidth]{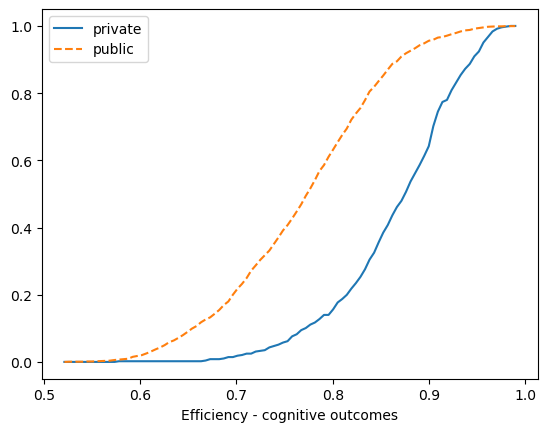}\label{figure2a}}
\subfloat[\footnotesize{First order SD-non cognitive outcomes}]{\includegraphics[width=.49\textwidth]{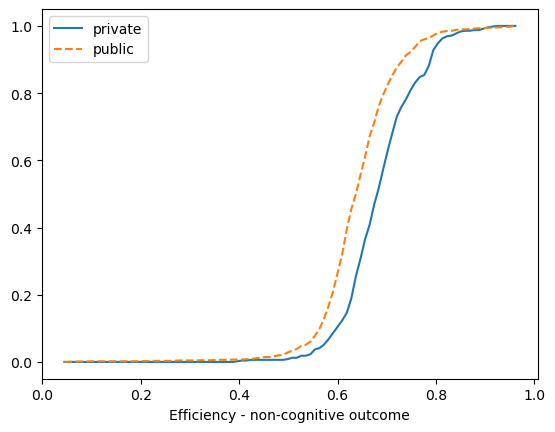}\label{figure2b}}
\\
\subfloat[\footnotesize{Second order SD-cognitive outcomes}]{\includegraphics[width=.49\textwidth]{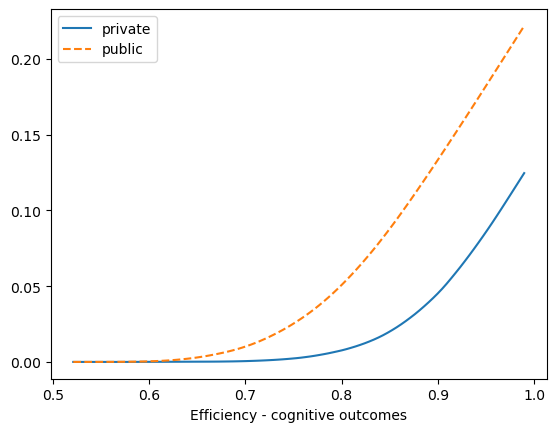}\label{figure2c}}
\subfloat[\footnotesize{Second order SD-non cognitive outcomes}]{\includegraphics[width=.49\textwidth]{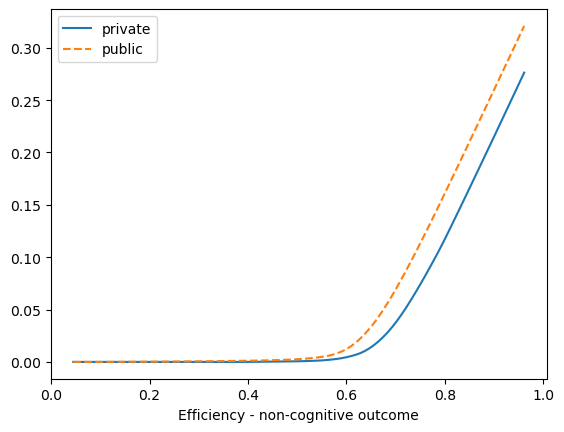}\label{figure2d}}
\caption{Cumulative density functions of efficiency scores by school type}
\label{figure2}
\end{figure}

Here, I formally compare the profile of school efficiency by school type by running stochastic dominance (SD) tests of efficiency scores. Results are included in Table \ref{table5} (and cumulative densities of efficiency scores of private and public schools are displayed in Figure \ref{figure2}). First order SD tests for both outcomes are shown in Table \ref{table5} (columns 1 and 3). Tests results indicate all hypotheses of the first SD order ($s$ = 1), equally for cognitive efficiency and non-cognitive efficiency specifications, are accepted (p-values = 1.0000 and 0.9450). Then globally --across the spectrum of efficiency scores estimated-- private schools are more efficient in the two efficiency fronts. The same can be seen when looking at the cumulative densities of private schools which are below the ones for public schools (Figure \ref{figure2a} and Figure \ref{figure2b}). Relating the test's findings to the economic domain, it means that an educational planner with an increasing utility function (higher efficiency leading to higher utility) would prefer the profile of efficiency of private schools. Similarly, the hypothesis of private schools second order dominating public schools for the two DEA models are accepted ($s$ = 2, columns 2 and 4 of Table \ref{table5}); which, in turn, it implies that an educational planner would prefer the lower between-schools dispersion of private schools (see: Figures \ref{figure2c} and Figure \ref{figure2d}).

\medskip
\begin{center}
  [Table \ref{table5} and Figure \ref{figure2} here]
\end{center}
\medskip

\subsection{Interpretable machine learning estimates}
\label{section52}

After finding out that private schools outperform public schools in both cognitive and non-cognitive efficiency in Section \ref{section51}, a follow-up empirical question is to examine whether the private-public school efficiency gap can be attributed to specific factors placing private schools in a better position to operate and to be able to produce more educational outputs with the same (or lower) levels of inputs. Here, I answer this question using IML techniques (i.e., the SHAP method; \citealp{lundberg17}), which allows to jointly evaluate the relevance of all students and schools covariates. This is carried out in two stages: (i) a global approach where estimates or SHAP values are the average impacts across all schools; and (ii) a local approach where the focus is on SHAP values for specific observations (or schools). Recall that this second stage uses a binary dependent variable for cognitive and non-cognitive efficiency (=1 if efficiency is above the regional mean, and 0 otherwise), and that mean SHAP values are presented as absolute values so that rankings of the global approach do not consider the direction of determinants' associations with efficiency but, instead, their absolute empirical relevance.

\subsubsection{Drivers of school efficiency--global explanations}
\label{section521}

\begin{figure}[ht!]
\centering
\subfloat[\footnotesize{Cognitive outcomes}]{\includegraphics[width=.99\textwidth]{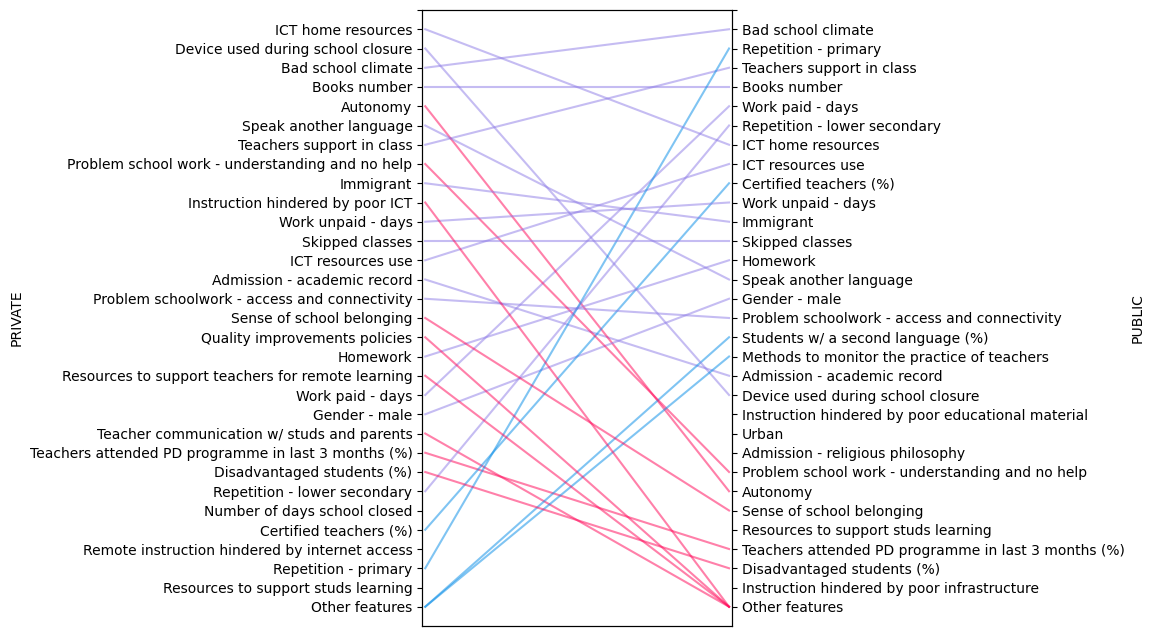}\label{figure3a}}
\\
\subfloat[\footnotesize{Non-cognitive outcomes}]{\includegraphics[width=.99\textwidth]{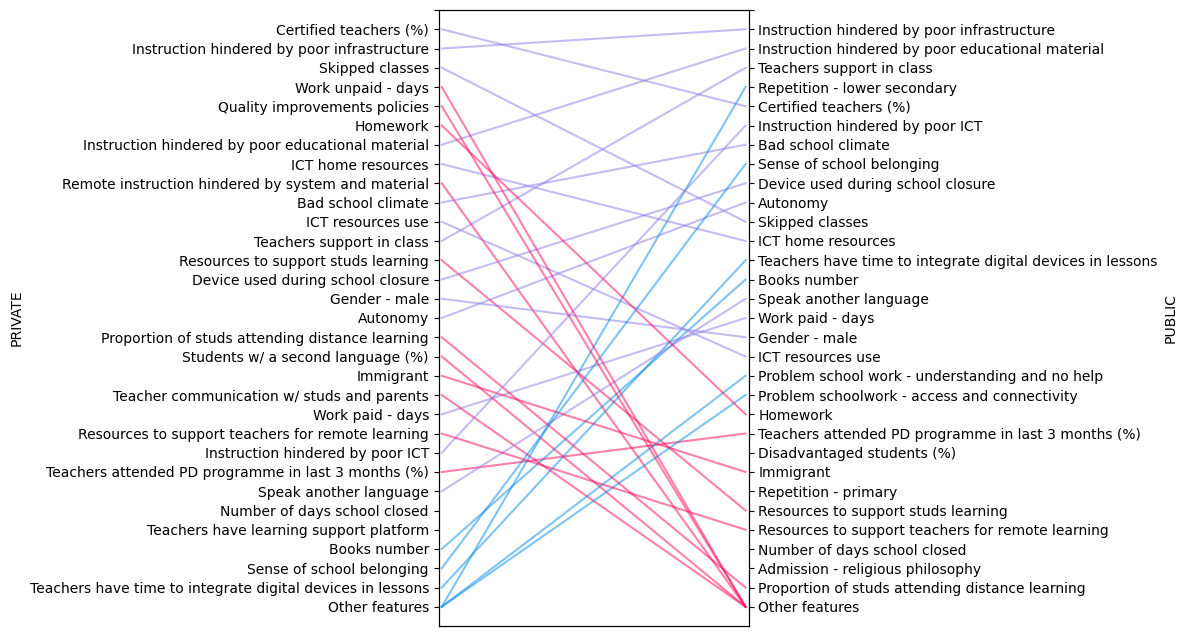}\label{figure3b}}
\caption{Comparison of determinants of school efficiency}
\label{figure3}
\medskip
\begin{minipage}{0.99\textwidth}
\footnotesize{
Notes: (1) Importance plots show ranked average absolute SHAP values per covariate (top 30 covariates). (2) SHAP values for these covariates are shown in Figure \ref{figureD1}.
}
\end{minipage}
\end{figure}

Out of the total list of student and school determinants of efficiency included in Table \ref{table2} (42 covariates in total), Figure \ref{figure3} showcases the top 30 leading covariates of GBT estimates in predicting the probability that schools efficiency falls into the top half of performing schools by ranking covariates importance based on SHAP values.\footnote{For completeness, SHAP values for these covariates are shown in the Appendix (Figure \ref{figureD1}).} First, Figure \ref{figure3a} displays the rank of determinants for the efficiency cognitive model, and for easier comparison estimates for the two models are shown along each other, with results for private schools' determinants shown on the left and those for public schools on the right. The plot shows that among private schools and public schools there is, overall, a lack of correspondence for the rank of environmental variables impacts on the chances of a school being top half of efficient schools ($\widehat{\theta}_{i}^{\text{yc}} > \overline{\theta}^{\text{yc}}$ or $\theta^{\text{binary, yc}}_{i}$ = 1). Top ranked covariates for the private school model are linked to resources (ICT resources at home, using a digital device during the pandemic, book stock at home) and variables related to schools' functioning (e.g., school climate, autonomy, teachers' support in class) and, by contrast, in the public school model, top environmental variables are --as well as (bad) school climate-- those related to students weak educational trajectories (i.e., repetition at primary and at secondary levels) and a student's disadvantage (such as engagement in paid work). This sharp contrast between top-ranked covariates is exemplified with the divergent ranks of repetition and work status covariates (second and fifth in public schools, and ranked 29 and 20 in private schools). Moreover, among the lower ranked students and schools characteristics, variables from the policy domain (e.g., quality improvements, remote learning teacher policies on resources and communication) are also relatively better ranked (and so have higher leverage effects on school efficiency) in private schools than in public schools.

\medskip
\begin{center}
  [Figure \ref{figure3} here]
\end{center}
\medskip

Second, the order of determinants for non-cognitive efficiency based on average SHAP (absolute) values are displayed in Figure \ref{figure3b}. The plot shows that, among the leading top-10 determinants (with either positive or negative influence) placing a school with a probability to be in the top half performing public schools, are instruction barriers (physical infrastructure, educational material and ICT) and, again, repetition (though at lower secondary level), teachers' variables (support in class and rate of certification), school climate and students' sense of belonging (to school).\footnote{Conversely, lower secondary repetition, instruction hindered by poor ICT, and sense of school belonging are well below in the ranking for private schools (falling into the `Other features' group), ranked below the position 30.} In the case of private schools, unpaid work, quality improvement policies and homework are some higher ranked determinants compared to public schools, and similar rankings are found to public schools for rate of certified teachers and schools' infrastructure.\footnote{It should be also noted that some COVID-19 pandemic covariates, such as the prevalence of students engaged in remote learning and communication of teachers with students and parents, are higher ranked in private schools.}

\subsubsection{Schools at extremes of SHAP index distribution. Experiment--local explanations for school efficiency determinants}
\label{section522}

In this section, I focus on local explanations of SHAP values which are given by specific schools, proceeding as follows. I calculate the mean sample contribution of SHAP values for each school and sort them out and then I select schools with the minimum and maximum contributions so that to profile the impact of covariates at extremes. Once school pairs are identified, I plot top determinants for each school, with plots showing the specific value each covariate takes. Given the differential on efficiency levels I found in Section \ref{section51} of the DEA analysis in favour of private schools, this local exercise comparison is between a private school with a highest index (DEA $\mapsto$ 1, a school on the efficiency frontier) and a public school with the lowest index (DEA $\mapsto$ 0, a school furthest away from the efficient frontier).

\begin{figure}[ht!]
\centering
\subfloat[\scriptsize{Private schools (cognitive outcomes). Profile high}]{\includegraphics[width=.49\textwidth]{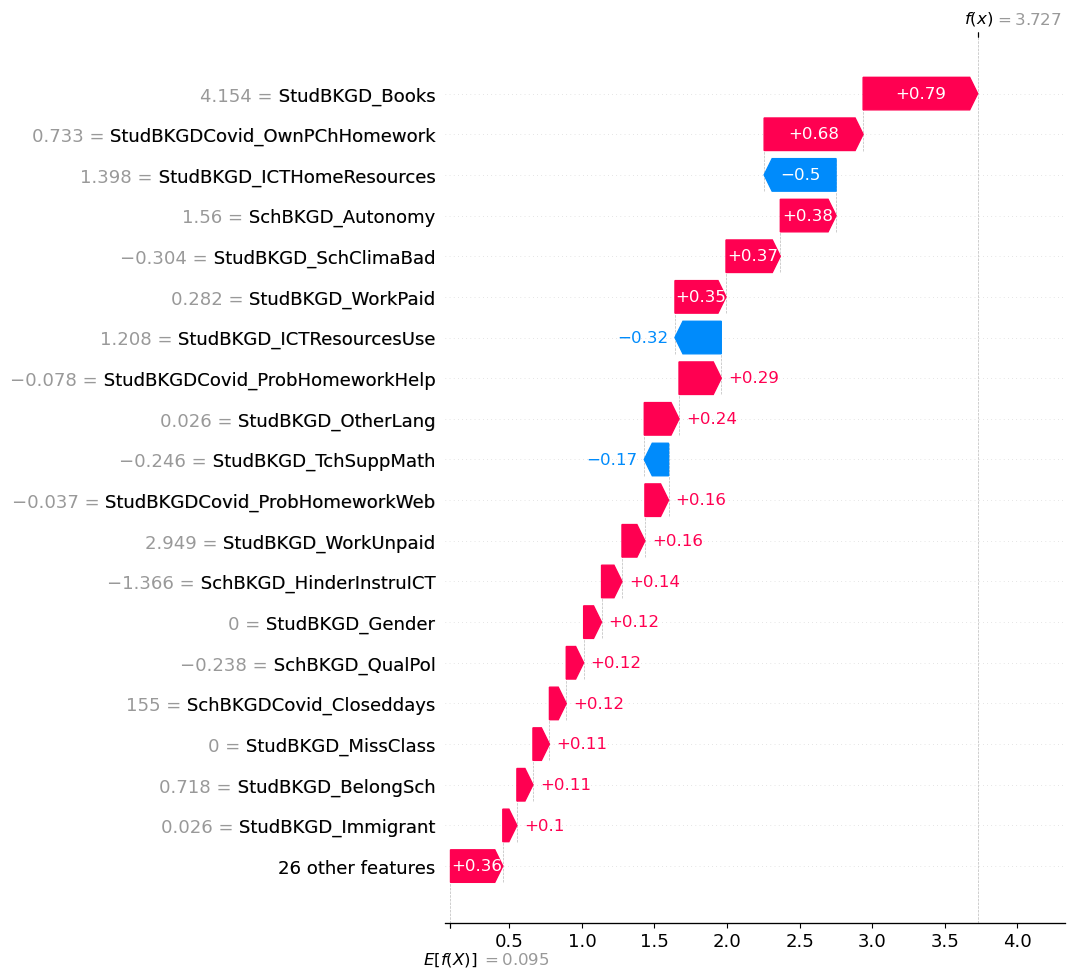}\label{figure4a}}
\subfloat[\scriptsize{Public schools (cognitive outcomes). Profile low}]{\includegraphics[width=.49\textwidth]{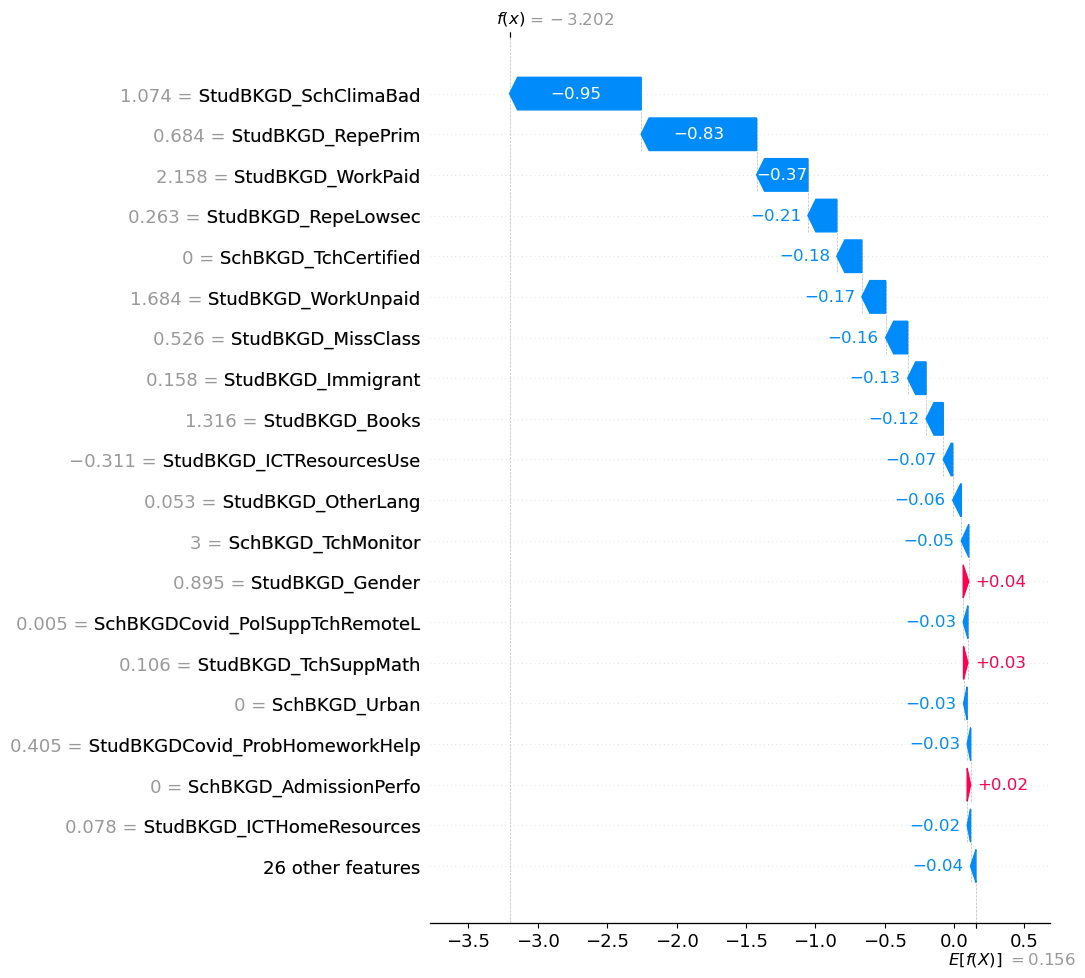}\label{figure4b}}
\\
\subfloat[\scriptsize{Private schools (non-cognitive outcomes). Profile high}]{\includegraphics[width=.49\textwidth]{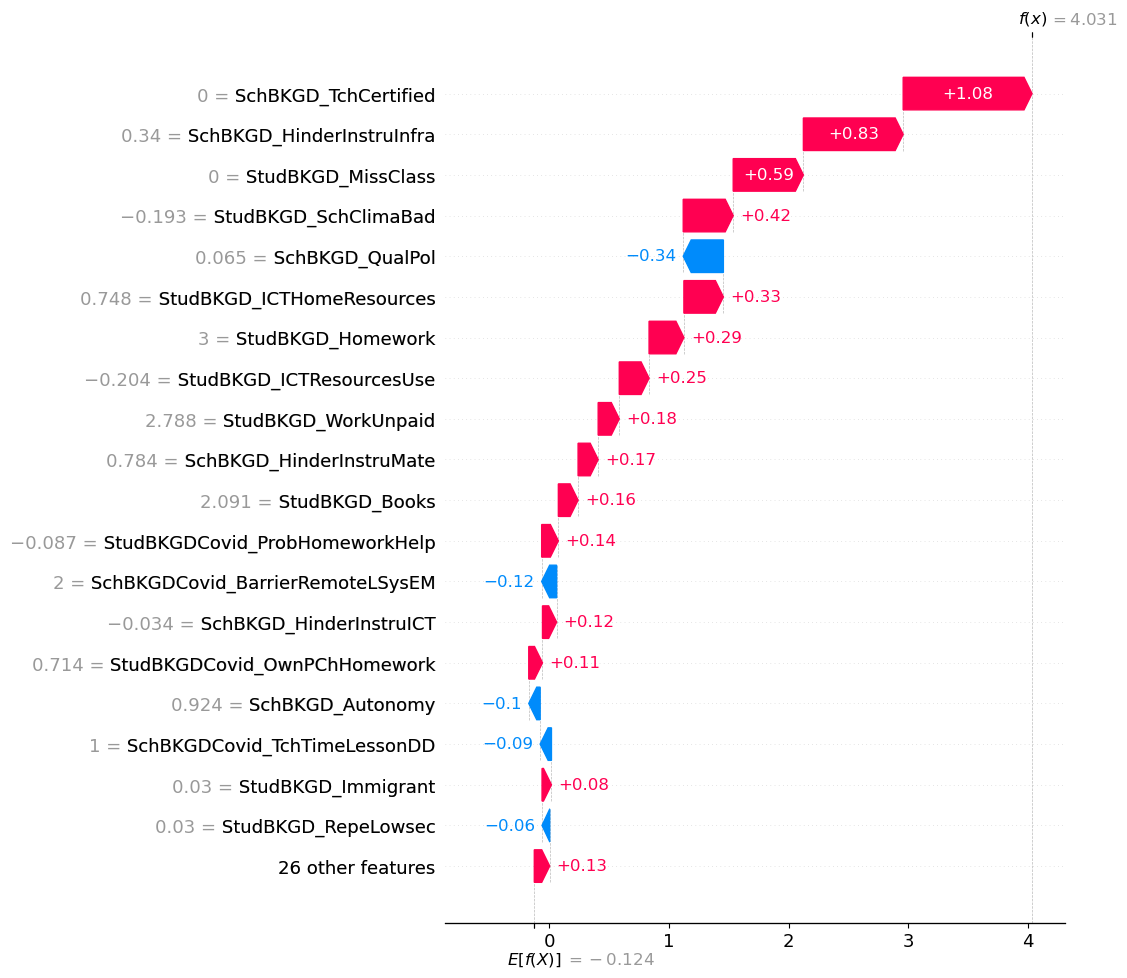}\label{figure4c}}
\subfloat[\scriptsize{Public schools (non-cognitive outcomes). Profile low}]{\includegraphics[width=.49\textwidth]{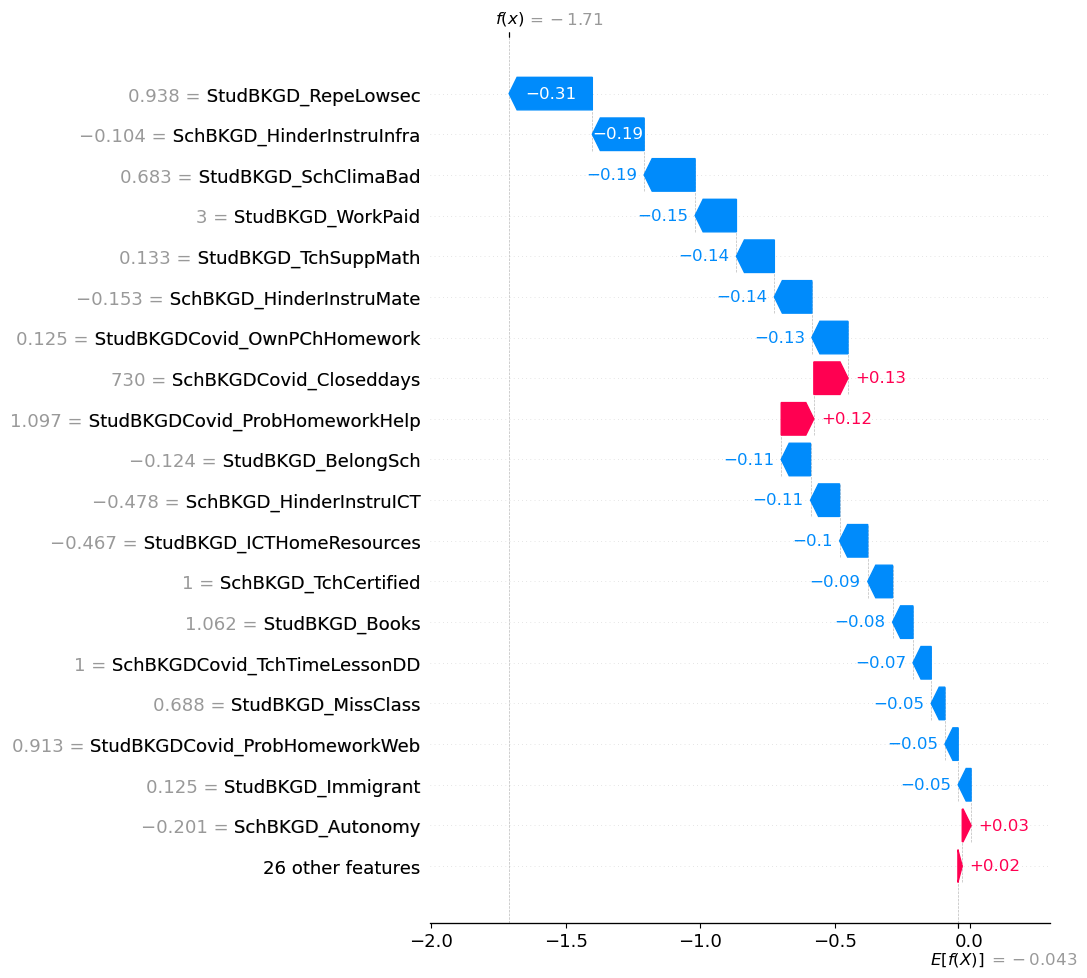}\label{figure4d}}
\caption{Covariates specific SHAP values for private schools with highest values contributions and public schools with lowest SHAP values contributions}
\label{figure4}
\medskip
\begin{minipage}{0.99\textwidth}
\footnotesize{
Notes: (1) See Table \ref{tableD1} for variables' acronyms, definitions and associated labels to values (in grey, left to covariate name).
}
\end{minipage}
\end{figure}

This exercise for the cognitive efficiency model, where I compare a private school with the largest chance to be highly efficient ($\Phi_{i,max}^{\text{private,yc}}$ = +3.63) and a public school with the lowest chance of being a top performing school ($\Phi_{i,min}^{\text{public,yc}}$ = -3.36), is shown in Figures \ref{figure4a} and \ref{figure4b}. A private school with the highest probability of being highly efficient (placing it on the efficiency frontier; Figure \ref{figure4a}) has, on average, students with a large number of books at home (between 101-200, $X$ = 4), a substantial rate of students with their own PC used for homework (rate = 73\%), it has larger autonomy (index = 1.398 falling into the top half group) and also a good school climate (StudBKGD\_SchClimaBad = -0.304), a low rate of student's work (0.282, working less than 1 day per week) and students with a second language (2.6\%), and their students do not have issues with the content of homework during COVID-19 (StudBKGDCovid\_ProbHomeworkHelp = -0.078). Similarly, though with lower SHAP values, this highly efficient private school has students who report not having issues with connectivity for homework during the pandemic (StudBKGDCovid\_ProbHomeworkWeb = -0.037) and it has no barriers for learning due to ICT (SchBKGD\_HinderInstruICT = -1.366) coupled with a lack of truancy and high sense of belonging to school (StudBKGD\_BelongSch = 0.718).

\medskip
\begin{center}
  [Figure \ref{figure4} here]
\end{center}
\medskip

On the opposite direction, a lowest efficient public school's profile has a poor school climate (index = 1.074), large rates of repetition (primary = 68.4\%, lower secondary = 26.3\%), no certified teachers, high intensity of students' work (e.g., StudBKGD\_WorkPaid = 2.158, working between 2-3 days weekly), significant prevalence of school truancy (StudBKGD\_MissClass = 52.6\%) and immigrant students (15.8\%)  with few books at home (=1.316, between 1 to 10 books) (Figure \ref{figure4b}). Additionally, other characteristics of this public school (though with smaller impacts) are its weak teachers' policy support (SchBKGDCovid\_PolSuppTchRemoteL = 0.005) and large issues of students to do their homework during the pandemic (StudBKGDCovid\_ProbHomeworkHelp = 0.405).

As regards to the non-cognitive efficiency schools' comparison (Figures \ref{figure4c} and \ref{figure4d}), leading drivers placing either private schools at the top of efficiency ($\Phi_{i,max}^{\text{private,ync}}$ = +4.15)  and public schools at the bottom of efficiency ($\Phi_{i,min}^{\text{public,ync}}$ = -1.67) are as follows. One the one hand, low soft-skills levels at a given input level is associated with a public school with very large rates of repetition (at lower secondary, of 93\%), very high intensity of students' paid work (3 days per week), poor school climate (= 0.683), reduced home ownership of PCs for remote learning's homework (12.5\%) and consequently with serious issues to carry out homework due to connectivity (StudBKGDCovid\_ProbHomeworkWeb = 0.913), a sizable period of shutdown (= 730 days, well above the 272 average days of closure of public schools), and a significant rate of school truancy (StudBKGD\_MissClass = 0.688). On the other hand, a private school which is vastly efficient in the production of soft-skills is a school without truancy and a decent students' engagement in homework in standard settings (=3, 1 to 2 hours per week), a good school climate (= 0.193), very high values both for autonomy and rates for student's PC availability for homework during the pandemic.


\section{Discussion and concluding remarks}
\label{section6}

In a challenging educational context post COVID-19 pandemic for the Latin American and the Caribbean (LAC) region, with diminishing learning levels and widening inequality (\citealp{arias23,bracco25,brunori24,fernandez24}) which has led to a deepening of segregation of education systems shifting sub-equilibriums of private schools and public schools and their functioning through governments' reallocation of resources, finding out what are the returns of educational investments is of crucial importance. On the back of this context --and the scarcity of investments for the LAC education sector-- knowing which is the degree of efficiency (and inefficiency levels) of private and public schools is vital so that policy makers can make reasonable adjustments through appropriate policy interventions. This paper offered some insights on this by estimating the degree of efficiency at secondary level (of two kinds: based on academic outputs and soft-skills) by school type and their determinants as a roadmap for the educational sector's future improvements efficiency-wise, so that schools can make the most of educational cognitive and non-cognitive outputs at given level of inputs employed. In particular, based on the latest 2022 PISA data for nine LAC countries comprising 2,034 schools, I obtained private and public schools' efficiency scores using Data Envelopment Analysis (DEA) and, then, to accommodate the large number of determinants (operating at the student's and school's levels) and to assess how they are ranked as drivers of school’s efficiency, I relied upon recent advancements on interpretable machine learning methods (IML) (\citealp{chen23,lundberg17}).

The paper contributed to the literature in at least three different aspects. First, to the best of my knowledge, this is the first post COVID19-pandemic comprehensive study for the LAC region estimating school efficiency and their determinants by school type, expanding earlier pre-pandemic country empirical evidence (\citealp{delprato21a}) to a regional level. Second, the paper innovates in the estimation of school efficiency domain by addressing non-cognitive efficiency given its relevance during the pandemic for academic engagement and performance (\citealp{caycho23,delprato25b,li25}). Third, the paper relies on a hybrid approach, combining DEA with IML methods, adding to the this very incipient and small methodological literature (\citealp{boubaker25,chen21}) and, for the first time, extending it towards the cross-country international education efficiency domain for the global south.

I find a superior performance of private schools in comparison to public schools, both on cognitive and non-cognitive domains, at a given level of inputs employed. The private-public schools efficiency gap is around 0.10 for cognitive efficiency (private school efficiency = 0.865, public school efficiency =0.768) and of 0.045 for non-cognitive (private school efficiency = 0.685, public school efficiency =0.640).  As shown by the stochastic dominance analysis of estimated efficiency frontiers by school type, the superior leverage mechanisms and technological linkage when transforming educational inputs into outputs of secondary schools from the private sector of the LAC region happens all across spectrum of school efficiency profiles, be it at low, at medium or at high schools’ efficiency brackets.

Still, there is a considerable scope to boost efficiency in the public sector at the current stock of educational inputs employed (by 24\% in the case of cognitive outcomes, for instance).  Policy-wise a reduction on cognitive efficiency’s heterogeneity between public schools (which is larger than in private schools: IQR difference of 0.034) is a way forward. In this homogenisation exercise of public schools efficiency, conducting benchmarking analyses within each country where one identifies practices and features of some schools that are highly efficient (which are able to make the most with their available resources) which then can be `transferred' to low-efficient public schools. In the same way, the lower gap on non-cognitive efficiency’s finding (half of the one found for cognitive outcomes) is an interesting platform for narrowing efficiency gaps by school type and, because of the positive correlation I found across cognitive and non-cognitive efficiency (which is larger in public schools than in private schools), improving public schools’ students personality traits can eventually lead towards a higher efficiency based on academic scores acting as a mechanism against contextual adversity (\citealp{windle11}) of the schools they attend. Enhancing students’ soft skills (be it on assertiveness, autonomy, emotional control, empathy, perseverance, etc.)  through financial support of policies and initiatives can, therefore, be a pathway to reduce school type’s efficiency gaps, particularly in situations of remote$/$online learning (\citealp{lagones24,zhoc22}).

As regards to IML’s local explanatory analysis of schools efficiency determinants by means of a comparison of SHAP values of high efficient private schools versus a low-efficient public school using cognitive outcomes, I find that a hypothetical private school with a highest probability of being highly efficient (located on the efficiency frontier) has better family inputs (such as larger numbers of books, ICT for homework, and hardly no students' engagement in paid-work) and superior school’s material and environmental working conditions (e.g., good connectivity for online work during the pandemic, no ICT barriers for learning, enhanced sense of school’s belonging and lack of students’ truancy). This is not unexpected. \cite{arias24a}'s report emphasises that, in the LAC region, private schools tended to better support learning continuity during school closures through remote instruction with their students more engaged in distance learning activities and private schools took more actions after the pandemic struck, with differences between public and private schools were larger than those between poorer and richer schools. Conversely, a public school in the other extreme at the bottom of school efficiency distribution has an array of disadvantages coming from its students population  (larger rates of repetition and truancy and high intensity of paid work, immigrant condition, low stock of books at home) and various factors negatively affecting remote learning of students (e.g., barriers to do their homework during the pandemic) and weak support concerning different teachers policies to adapt to remote learning.


\newpage
\printendnotes


\clearpage
\setlength\bibsep{0pt}
\bibliographystyle{elsarticle-harv}
\bibliography{references}


\newpage

\newpage

\newpage
\clearpage
\begin{table}[htbp!]
\begin{threeparttable}
\caption{Summary statistics. DEA specification: outputs and inputs}
\begin{tabular}{llllllll}
\hline
 & \multicolumn{2}{c}{Public schools} &  & \multicolumn{2}{c}{Private schools} &  & Difference \\
   \cmidrule{2-3} \cmidrule{5-6}
 & Mean & SD &  & Mean & SD & & \\
 \hline
 & (1) & (2) &  & (3) & (4) & & (5) \\
\textit{i. Outputs – cognitive} &  &  &  &  &  &  \\
Math & 367.07 & 42.94 &  & 426.36 & 51.82 & & 59.29 \\
Reading & 388.27 & 56.49 &  & 458.02 & 58.49 & & 69.75 \\
Science & 388.65 & 49.70 &  & 456.25 & 56.75 & & 67.60 \\
\textit{ii. Outputs – non-cognitive} &  &  &  &  &  &  \\
Soft skills index & 0.025 & 0.380 &  & 0.095 & 0.320 & & 0.07 \\
 &  &  &  &  &  &  \\
\textit{iii. Inputs} &  &  &  &  &  &  \\
Family SES & -1.221 & 0.614 &  & -0.101 & 0.715 &  & 1.12 \\
School infrastructure - physical & -0.184 & 0.962 &  & 0.652 & 0.664 & & 0.84 \\
School infrastructure - educational & -0.140 & 0.896 &  & 0.515 & 0.685 & & 0.65 \\
Student teacher ratio (inverse) & 0.084 & 0.121 & & 0.086 & 0.099 &  & \textbf{0.00} \\
 &  &  &  &  &  &  \\
N & \multicolumn{2}{c}{1,548} &  & \multicolumn{2}{c}{486} &  &  \\
\hline
\end{tabular}
\label{table1}
\begin{tablenotes}[para,flushleft]
\medskip
\footnotesize{
(1) Differences lacking statistically significance at 10\% are shown in bold.
}
\end{tablenotes}
\end{threeparttable}
\end{table}

\begin{table}[htbp!]
\begin{threeparttable}
\caption{Summary statistics by school type of efficiency determinants}
\begin{tabular}{llllllll}
\hline
 & \multicolumn{2}{c}{Public schools} &  & \multicolumn{2}{c}{Private schools} &  & Difference \\
   \cmidrule{2-3} \cmidrule{5-6}
 & Mean & SD &  & Mean & SD & & \\
 \hline
 & (1) & (2) &  & (3) & (4) & & (5) \\
\textit{Panel A - students variables} &  &  &  &  &  &  \\
Gender - male & 0.511 & 0.172 &  & 0.489 & 0.174 & & -0.02 \\
Homework & 3.323 & 1.109 &  & 3.412 & 0.874 & & \textbf{0.09} \\
Books number & 1.455 & 0.529 &  & 2.233 & 0.769 & & 0.78 \\
Speak another language & 0.043 & 0.108 &  & 0.029 & 0.092 & & -0.01 \\
Immigrant  & 0.049 & 0.082 &  & 0.044 & 0.081 & & \textbf{-0.01} \\
Work unpaid – days & 3.192 & 0.828 &  & 2.869 & 0.819 & & -0.32 \\
Work paid – days & 1.256 & 0.830 &  & 0.892 & 0.660 & & -0.36 \\
Repetition – primary & 0.183 & 0.254 &  & 0.072 & 0.165 & & -0.11 \\
Repetition – lower secondary & 0.146 & 0.228 &  & 0.048 & 0.127 &  & -0.10 \\
Skipped classes & 0.276 & 0.212 &  & 0.203 & 0.205 & & -0.07 \\
Sense of school belonging  & 0.013 & 0.368 &  & 0.026 & 0.398 & & \textbf{0.01} \\
Bad school climate, index & 0.019 & 0.487 &  & -0.091 & 0.626 & & -0.11 \\
Teachers support in class, index  & 0.221 & 0.424 &  & 0.111 & 0.454 & &  -0.11 \\
ICT home resources, index & -0.252 & 0.473 &  & 0.346 & 0.492 & & 0.60 \\
ICT resources use, index & -0.179 & 0.388 &  & 0.044 & 0.445 & & 0.22 \\
Device used during school closure & 0.245 & 0.203 &  & 0.511 & 0.251 & & 0.27 \\
Problem schoolwork – access and connectivity, index & 0.023 & 0.505 &  & 0.019 & 0.400 & & \textbf{0.00} \\
Problem schoolwork – understanding and no help, index & 0.037 & 0.478 & & 0.001 & 0.422 & & \textbf{-0.04} \\
\textit{Panel B - school variables} &  &  &  &  &  &  \\
Urban & 0.646 & 0.478 &  & 0.854 & 0.354 & & 0.21 \\
Competition & 0.742 & 0.438 &  & 0.959 & 0.199 & & 0.22 \\
Admission - academic record & 0.375 & 0.484 &  & 0.576 & 0.495 & & 0.20 \\
Admission – area of residence & 0.503 & 0.500 &  & 0.286 & 0.452 & & -0.22 \\
Admission - religious philosophy & 0.234 & 0.424 &  & 0.558 & 0.497 & & 0.32 \\
Teachers attended PD programme in last 3 months (\%) & 0.419 & 0.347 &  & 0.435 & 0.371 & & \textbf{0.02} \\
Certified teachers (\%) & 0.630 & 0.437 &  & 0.490 & 0.439 & & -0.14 \\
Methods to monitor the practice of teachers & 2.471 & 1.133 &  & 2.673 & 0.940 & & 0.20 \\
Social and emotional skills policy & 0.727 & 0.446 &  & 0.827 & 0.378 & & 0.10 \\
Quality improvements policies, index & 0.252 & 0.530 &  & 0.295 & 0.451 & & \textbf{0.04} \\
Autonomy, index & -0.344 & 0.850 &  & 1.042 & 0.589 & & 1.39 \\
Instruction hindered by poor educational material, index & 0.220 & 0.995 &  & -0.563 & 0.763 & & -0.78 \\
Instruction hindered by poor infrastructure, index & 0.217 & 1.005 &  & -0.647 & 0.645 & & -0.86 \\
Instruction hindered by poor ICT & 0.249 & 0.936 &  & -0.669 & 0.847 & & -0.92 \\
Students w/ a second language (\%) & 0.071 & 0.219 &  & 0.040 & 0.165 & & -0.03 \\
Disadvantaged students (\%) & 0.444 & 0.333 &  & 0.185 & 0.288 & & -0.26 \\
\textit{Panel C - COVID school variables} &  &  &  &  &  &  \\
Resources to support studs learning & 0.104 & 0.481 &  & 0.148 & 0.583 & & 0.04 \\
Teacher communication w/ studs and parents & 0.143 & 0.639 &  & 0.212 & 0.822 & & 0.07 \\
Resources to support teachers for remote learning & 0.063 & 0.291 &  & 0.104 & 0.394 & & 0.04 \\
Proportion of studs attending distance learning & 8.108 & 2.001 &  & 9.791 & 1.584 & & 1.68 \\
Teachers have skills for instruction & 0.812 & 0.391 &  & 0.915 & 0.278 & & 0.10 \\
Teachers have time to integrate digital devices in lessons  & 0.534 & 0.499 &  & 0.661 & 0.474 & & 0.13 \\
Teachers have learning support platform & 0.538 & 0.499 &  & 0.784 & 0.412 & & 0.25 \\
Remote instruction hindered by digital devices access & 5.455 & 1.473 &  & 3.852 & 1.390 & & -1.60 \\
Remote instruction hindered by internet access & 5.590 & 1.397 &  & 3.909 & 1.408 & & -1.68 \\
Remote instruction hindered by system and material & 4.580 & 1.792 & & 3.010 & 1.427 & & -1.57 \\
Number of days school closed & 272 & 163 &  & 232 & 152 & & -39 \\
\hline
\end{tabular}
\label{table2}
\begin{tablenotes}[para,flushleft]
\medskip
  \footnotesize{
(1) Differences lacking statistically significance at 10\% are shown in bold.
}
\end{tablenotes}
\end{threeparttable}
\end{table}

\newpage
\begin{table}[htbp!]
\begin{threeparttable}
\caption{Efficiency scores estimates. Output: cognitive, learning scores}
\begin{tabular}{lllllll}
\hline
 & TE  & TEBC & TEBC  & TEBC  & Bootstrap & N \\
 &  &  & lower bound & upper bound & performance & \\
 \hline
 & (1) & (2) & (3) & (4) & (5) & (6) \\
\textit{Panel A - Whole sample } &  &  &  &  &  \\
Public schools & 0.800 & 0.768 & 0.790 & 0.744 & 29.70 & 1,584 \\
Private schools & 0.891 & 0.865 & 0.887 & 0.840 & 21.62 & 486 \\
&  &  &  &  &  &  \\
TEBC scores & Mean & SD & IQR & Min & Max & N \\
\hline
\textit{Panel B – Public schools} &  &  &  &  &  \\
Whole sample & 0.768 & 0.081 & 0.117 & 0.521 & 0.990 & 1,584 \\
Argentina & 0.764 & 0.073 & 0.098 & 0.586 & 0.961 & 245 \\
Brazil & 0.779 & 0.085 & 0.108 & 0.574 & 0.990 & 329 \\
Chile & 0.786 & 0.075 & 0.098 & 0.607 & 0.946 & 59 \\
Colombia & 0.786 & 0.069 & 0.101 & 0.629 & 0.953 & 164 \\
Dominican Rep. & 0.664 & 0.048 & 0.069 & 0.545 & 0.835 & 122 \\
Mexico & 0.782 & 0.068 & 0.088 & 0.579 & 0.947 & 217 \\
Panama & 0.708 & 0.077 & 0.113 & 0.521 & 0.890 & 97 \\
Peru & 0.789 & 0.068 & 0.089 & 0.559 & 0.976 & 206 \\
Uruguay & 0.795 & 0.079 & 0.110 & 0.613 & 0.959 & 145 \\
 &  &  &  &  &  &  \\
\textit{Panel C – Private schools} &  &  &  &  &  \\
Whole sample & 0.865 & 0.066 & 0.083 & 0.573 & 0.983 & 486 \\
Argentina & 0.845 & 0.057 & 0.065 & 0.688 & 0.966 & 113 \\
Brazil & 0.891 & 0.067 & 0.078 & 0.573 & 0.975 & 49 \\
Chile & 0.889 & 0.053 & 0.066 & 0.671 & 0.978 & 104 \\
Colombia & 0.874 & 0.061 & 0.079 & 0.717 & 0.974 & 42 \\
Dominican Rep. & 0.761 & 0.049 & 0.065 & 0.672 & 0.899 & 26 \\
Mexico & 0.861 & 0.063 & 0.086 & 0.710 & 0.961 & 31 \\
Panama & 0.819 & 0.067 & 0.116 & 0.701 & 0.931 & 23 \\
Peru & 0.883 & 0.058 & 0.078 & 0.664 & 0.983 & 75 \\
Uruguay & 0.895 & 0.039 & 0.069 & 0.825 & 0.968 & 23 \\
    \hline
    \end{tabular}%
  \label{table3}%
\begin{tablenotes}[para,flushleft]
  \medskip
  \footnotesize{
Notes: (1) Technical efficiency (TE) scores are radial measure of technical efficiency, and they are derived using variable returns to scale (VRS) and homogenous bootstrap (for bias corrected version). Results from these two tests are available from the author upon request. (2) TE: technical efficiency scores, output orientated (inverted); TEBC: technical efficiency scores bias corrected using 2,000 bootstrap repetitions. (3) Lower and upper bootstrap confidence interval at 95\% level. (4) Bootstrap performance is given by three times the ratio of bias squared to variance for radial measures of technical efficiency; the statistic is large so it indicates the appropriateness of bias correction (smoothed homogenous bootstrap). (5) Estimation commands used are: \texttt{teradial} and \texttt{teradialbc} (see: \citealt{badunenko16}). (6) All differences between private schools and public schools, both for the whole sample and for country samples, are statistically significant at 10\%.
}
\end{tablenotes}
\end{threeparttable}
\end{table}%

\begin{table}[htbp!]
\begin{threeparttable}
\caption{Efficiency scores estimates. Output: non-cognitive}
\begin{tabular}{lllllll}
\hline
 & TE  & TEBC & TEBC  & TEBC  & Bootstrap & N \\
 &  &  & lower bound & upper bound & performance & \\
 \hline
 & (1) & (2) & (3) & (4) & (5) & (6) \\
\textit{Panel A - Whole sample } &  &  &  &  &  \\
Public schools & 0.679 & 0.640 & 0.667 & 0.606 & 21.99 & 1,584 \\
Private schools & 0.726 & 0.685 & 0.715 & 0.646 & 19.42 & 486 \\
&  &  &  &  &  &  \\
TEBC scores & Mean & SD & IQR & Min & Max & N \\
\hline
\textit{Panel B – Public schools} &  &  &  &  &  \\
Whole sample & 0.640 & 0.080 & 0.085 & 0.045 & 0.961 & 1,584 \\
Argentina & 0.639 & 0.073 & 0.076 & 0.197 & 0.856 & 245 \\
Brazil & 0.624 & 0.074 & 0.082 & 0.344 & 0.946 & 329 \\
Chile & \textbf{0.725} & 0.101 & 0.118 & 0.502 & 0.955 & 59 \\
Colombia & 0.630 & 0.063 & 0.067 & 0.411 & 0.821 & 164 \\
Dominican Rep. & 0.641 & 0.091 & 0.095 & 0.069 & 0.888 & 122 \\
Mexico & \textbf{0.646} & 0.086 & 0.083 & 0.092 & 0.961 & 217 \\
Panama & \textbf{0.631} & 0.108 & 0.074 & 0.045 & 0.961 & 97 \\
Peru & 0.646 & 0.065 & 0.090 & 0.428 & 0.809 & 206 \\
Uruguay & \textbf{0.648} & 0.066 & 0.064 & 0.213 & 0.872 & 145 \\
 &  &  &  &  &  \\
\textit{Panel C – Private schools} &  &  &  &  &  \\
Whole sample & 0.685 & 0.075 & 0.090 & 0.389 & 0.918 & 486 \\
Argentina & 0.669 & 0.066 & 0.080 & 0.541 & 0.918 & 113 \\
Brazil & 0.670 & 0.066 & 0.065 & 0.389 & 0.840 & 49 \\
Chile & \textbf{0.732} & 0.069 & 0.097 & 0.517 & 0.911 & 104 \\
Colombia & 0.678 & 0.065 & 0.079 & 0.545 & 0.892 & 42 \\
Dominican Rep. & 0.711 & 0.070 & 0.109 & 0.546 & 0.843 & 26 \\
Mexico & \textbf{0.664} & 0.095 & 0.115 & 0.402 & 0.872 & 31 \\
Panama & \textbf{0.667} & 0.077 & 0.088 & 0.506 & 0.856 & 23 \\
Peru & 0.667 & 0.082 & 0.090 & 0.415 & 0.844 & 75 \\
Uruguay & \textbf{0.666} & 0.036 & 0.046 & 0.573 & 0.742 & 23 \\
    \hline
    \end{tabular}%
  \label{table4}%
\begin{tablenotes}[para,flushleft]
  \medskip
  \footnotesize{
Notes: (1) See notes of Table \ref{table2}. (2) Differences lacking statistically significance at 10\% are shown in bold.
}
\end{tablenotes}
\end{threeparttable}
\end{table}%

\begin{table}[htbp!]
\begin{threeparttable}
\caption{Stochastic dominance tests by school type for efficiency scores}
\begin{tabular}{llllll}
\hline
\textit{} & \multicolumn{2}{c}{Cognitive outcomes} &   & \multicolumn{2}{c}{Non-cognitive outcome} \\
 \cmidrule{2-3} \cmidrule{5-6}
H0 & \multicolumn{2}{c}{$\widehat{\theta}^{\text{private}} \preceq_{s} \widehat{\theta}^{\text{public}}$} &    & \multicolumn{2}{c}{$\widehat{\theta}^{\text{private}} \preceq_{s} \widehat{\theta}^{\text{public}}$} \\
\hline
\textit{s} & 1 & 2 &  & 1 & 2 \\
 & (1) & (2) &  & (3) & (4) \\
\textit{Results}: &  &  &  &   \\
Test statistic & 0.0122 & 0.0000 &  & 0.0853 & 0.0000 \\
Critical-value  & 1.0536 & 0.1043 &  & 1.1078 & 0.1197 \\
P-value & 1.0000 & 1.0000 &  & 0.9450 & 1.0000 \\
    \hline
    \end{tabular}%
  \label{table5}%
\begin{tablenotes}[para,flushleft]
  \medskip
  \footnotesize{
Notes: (1) First and second dominance tests results are obtained using the Python package \texttt{PySDTest} (\citealp{lee24}).
}
\end{tablenotes}
\end{threeparttable}
\end{table}%


\appendix


\renewcommand{\thesubsection}{\Alph{subsection}}
\setcounter{figure}{0}
\renewcommand{\thefigure}{A\arabic{figure}}
\setcounter{table}{0}
\renewcommand{\thetable}{A\arabic{table}}

\begin{table}[htbp!]
\caption{Comparison of ML models – parameters search}
\begin{tabular}{llll}
\hline
Grid search  & Chosen parameters & AUROC score & AUPRC score \\
\hline
 & (1) & (2) & (3) \\
\textit{(i). Logit} &  &  \\
Penalty (L1, L2) &  &  \\
C = (0.1, 1, 10) &  &  \\
Cognitive – public schools & L1, 0.1 & 0.797 & 0.811 \\
Cognitive – private schools & L2, 0.1 & 0.727 & 0.792 \\
Non-cognitive – public schools & L1, 0.1 & 0.661 & 0.606 \\
Non-cognitive – private schools & L1, 0.1 & 0.548 & 0.636 \\
 &  &  \\
\textit{(ii). Neural networks} &  &  \\
Hidden layer sizes &  &  \\
\{200\}, \{100, 100\}, \{200, 100, 50\} &  &  \\
Activation (relu, tanh, logistic) &  &  \\
Cognitive – public schools & \{200\}, logistic & 0.810 & 0.793 \\
Cognitive – private schools & \{200\}, relu & 0.678 & 0.721 \\
Non-cognitive – public schools & \{200\}, logistic & 0.632 & 0.573 \\
Non-cognitive – private schools & \{200, 100, 50\}, relu & 0.584 & 0.613 \\
 &  &  \\
\textit{(iii). Gradient boosted trees } &  &  \\
number of estimators &  &  \\
(100, 500, 1000, 5000) &  &  \\
subsample ratio (0.5, 0.7, 0.9) &  &  \\
trees of max depth (3, 5, 7, 9) &  &  \\
learning rate (0.001, 0.01, 0.1) &  &  \\
Cognitive – public schools & 500, 0.7 3, 0.01    & \textbf{0.836} & \textbf{0.834} \\
Cognitive – private schools & 100, 0.7, 5, 0.1 & \textbf{0.769} & \textbf{0.855} \\
Non-cognitive – public schools & 5000, 0.5, 3, 0.001 & \textbf{0.738} & \textbf{0.691} \\
Non-cognitive – private schools & 1000, 0.9, 5, 0.01 & \textbf{0.639} & \textbf{0.730} \\
\hline
\end{tabular}
\label{tableA1}
\end{table}


\newpage
\renewcommand{\thesubsection}{\Alph{subsection}}
\setcounter{figure}{0}
\renewcommand{\thefigure}{B\arabic{figure}}
\setcounter{table}{0}
\renewcommand{\thetable}{B\arabic{table}}

\begin{figure}[h!]
\centering
\subfloat[\footnotesize{Cognitive outcomes}]{\includegraphics[width=0.99\textwidth]{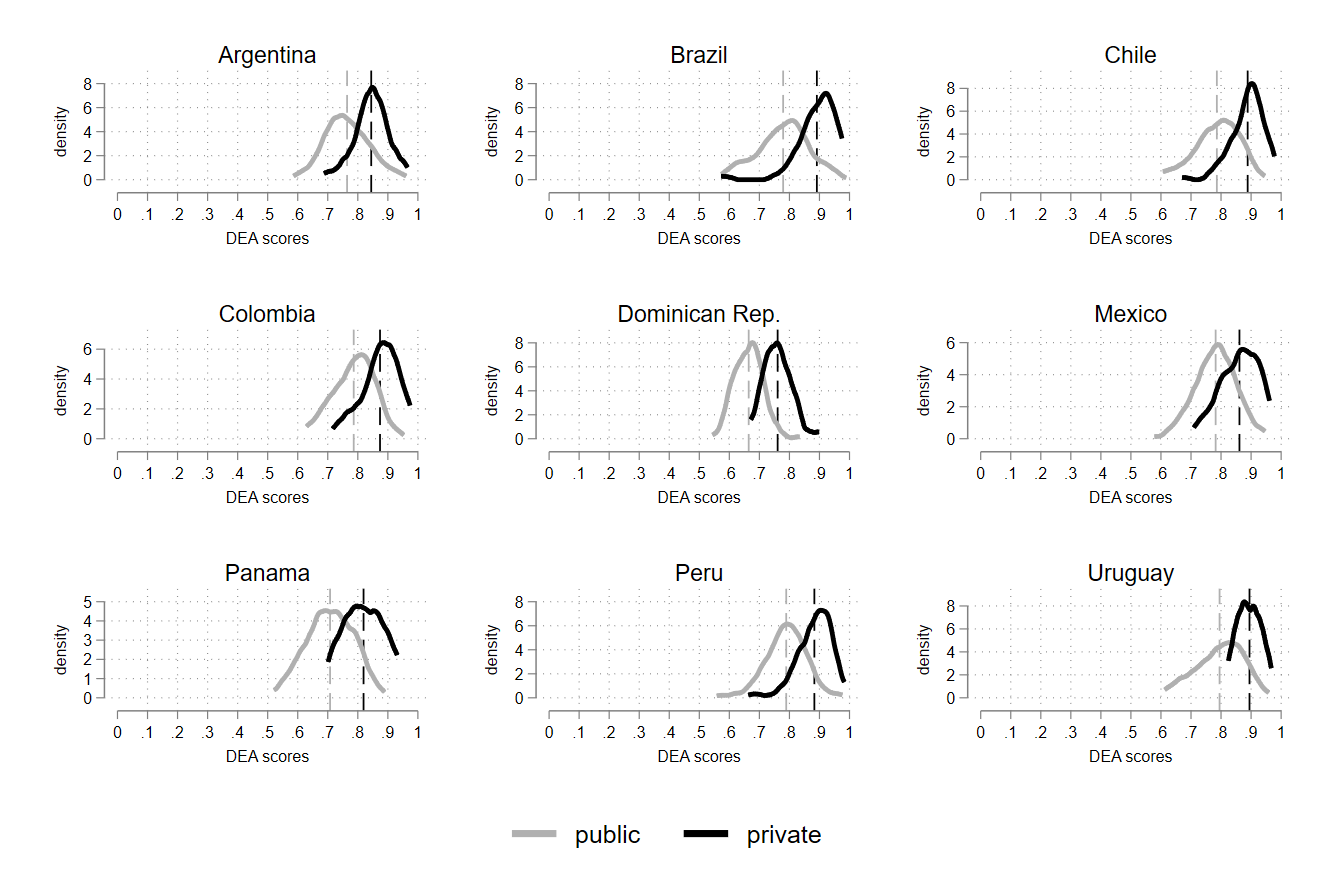}\label{figureB1a}}
\\
\subfloat[\footnotesize{Non-cognitive outcomes}]{\includegraphics[width=0.99\textwidth]{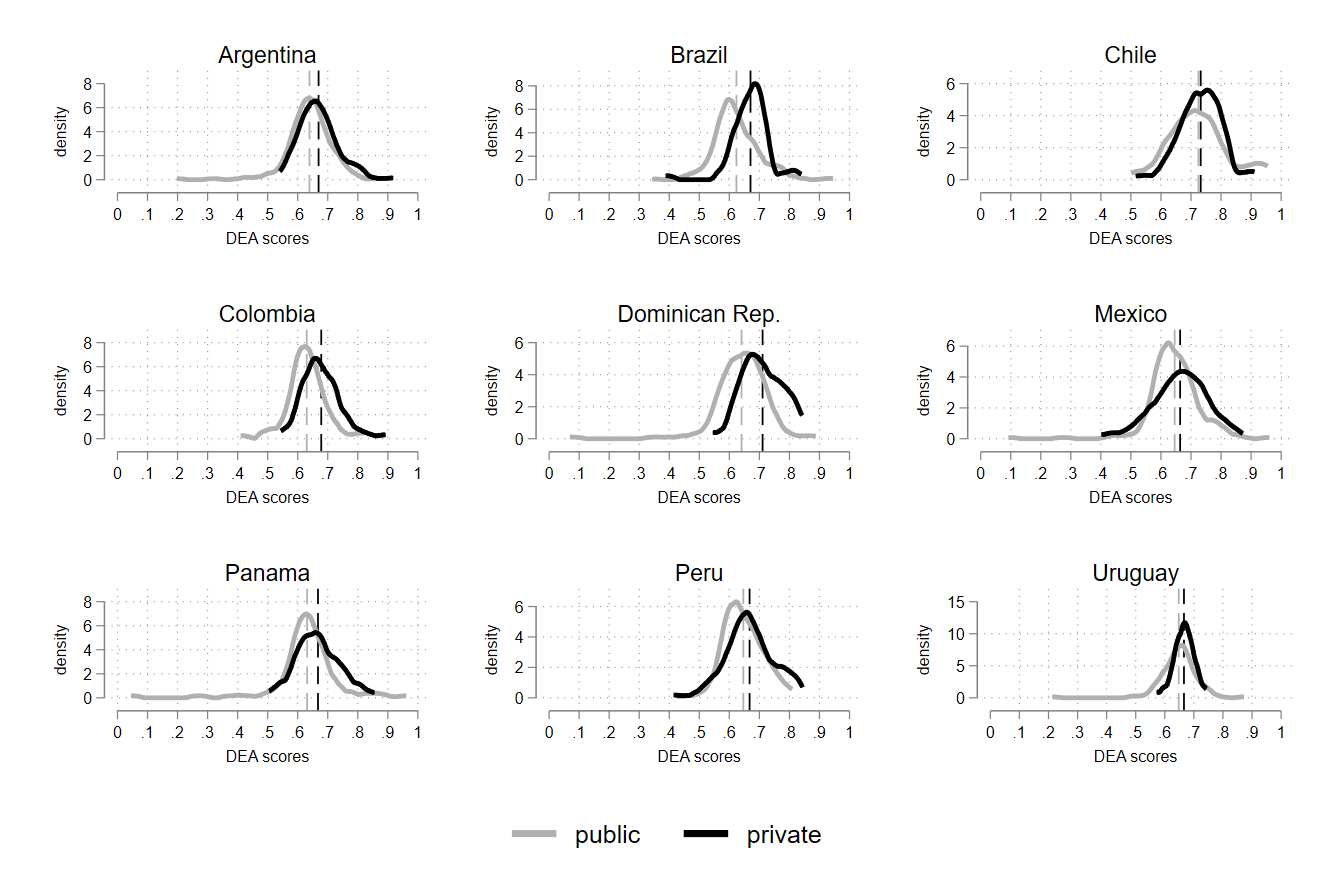}\label{figureB1b}}
\caption{Country densities of DEA scores by school type}
\label{figureB1}
\end{figure}


\clearpage
\newpage
\renewcommand{\thesubsection}{\Alph{subsection}}
\setcounter{figure}{0}
\renewcommand{\thefigure}{C\arabic{figure}}
\setcounter{table}{0}
\renewcommand{\thetable}{C\arabic{table}}
\subsection*{Appendix C—Robustness analysis}\label{appendixC}

The presence of outliers —stemming from wrong selection of inputs$/$ouputs, measurement error, heterogenous DMUs— can affect the performance of DEA's estimates (\citealp{boyd16,khezrimotlagh15}). The $\alpha$-order efficiency method of \cite{aragon05} is one of leading methods to address this issue identifying outliers by choosing a percentile on the efficiency distribution based on conditional quantiles of the distribution. Here, I assess the robustness of the DEA results (Section \ref{section51}) after accounting for the likelihood of school's outliers. To detect outliers, I follow the two-stage process of \cite{tauchmann12}. These two stages are: (i) to use the $\alpha$-order method for outliers detection which relies on the idea that the share of super-efficient DMUs should be decreasing, looking at a discontinuity point above which super-efficient DMUs are more prone to happen; and (ii) linking this discontinuity to $\alpha$, exclude those DMUs (school) and estimate the DEA model again with this smaller sample and comparing it to the original results.

\begin{figure}[ht!]
\centering
\includegraphics[width=0.99\textwidth]{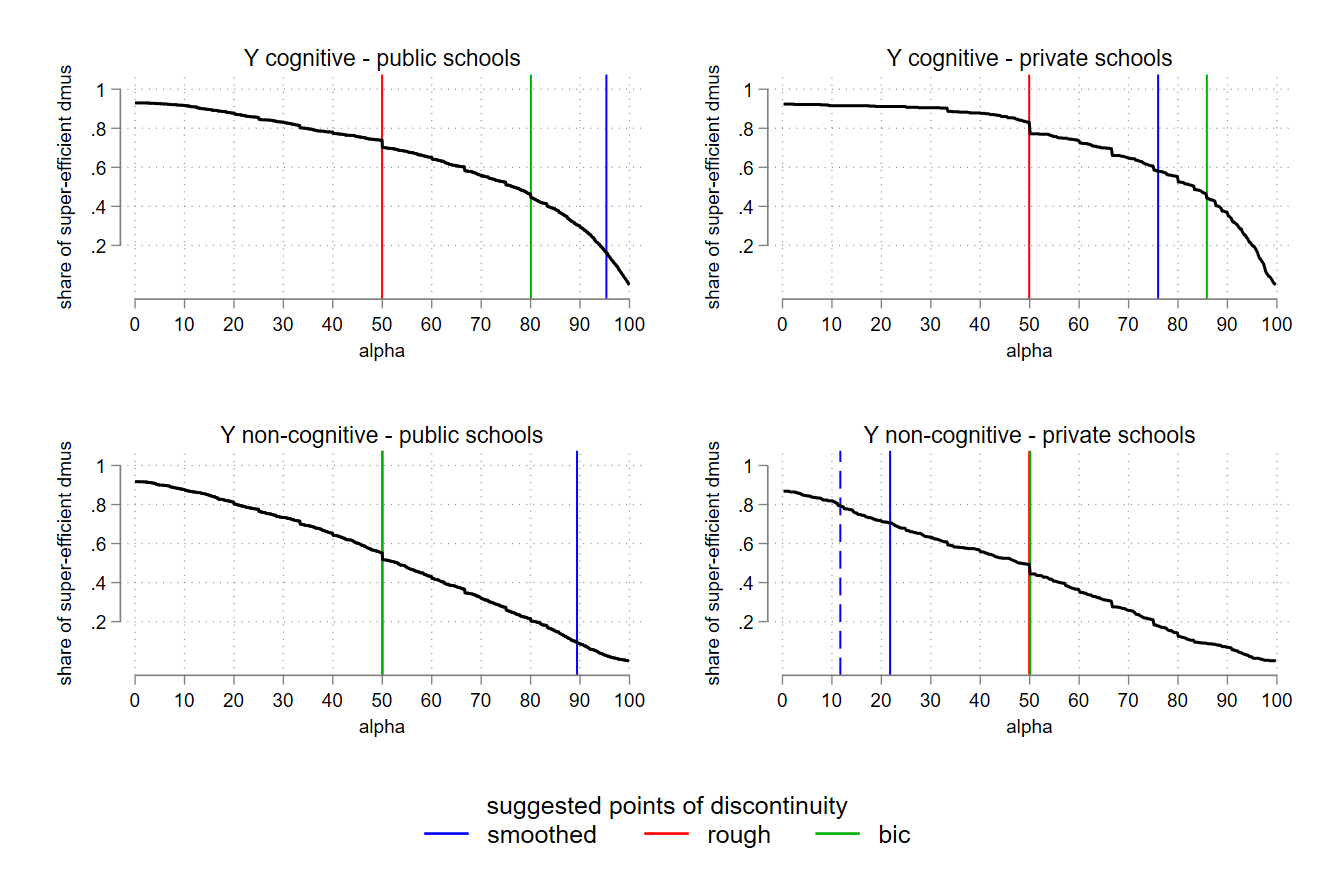}
\caption{Outliers detection based on order-$\alpha$ efficiency analysis. Share of supper-efficient schools}
\label{figureC1}
\medskip
\begin{minipage}{0.99\textwidth}
\footnotesize{
Notes: (1) Share of supper-efficient schools is for DEA's cognitive outcomes 16.48\% (public schools; order-$\alpha$ = 95.33) and 58.02\% (private schools; order-$\alpha$ = 75.98). (2) For DEA's non-cognitive outcome the rate of supper-efficient schools is 9.41\% (public schools; order-$\alpha$ = 89.39) and 70.58\%
(private schools; order-$\alpha$ = 21.77). (3) Estimation command used: \texttt{oaoutlier} (\citealp{tauchmann12}).
}
\end{minipage}
\end{figure}

Figure \ref{figureC1} shows the results of the four DEA models for the first stage. The plot indicates that points of discontinuities for the DEA cognitive outcomes models happen at values of $\alpha$ of 95 (public schools) and 58 (private schools) with a share of supper-efficient schools of around 16\% and 58\%, respectively; meanwhile for the non-cognitive outcome models' estimates are for public schools (order-$\alpha$ = 89, 9\% supper-efficient schools) and for private schools (order-$\alpha$ = 21, 70\% supper-efficient schools). Therefore, the presence of outliers seems to be more a concern for the private school sample (and particularly for the model using the non-cognitive outputs). After discarding these supper-efficient schools, new estimates of efficiency are obtained and proportional differences on estimates with the original (full) sample are shown in Table \ref{tableC1} (the second stage). As a whole, DEA estimates are not as different and they are somehow aligned with those based on the full sample of schools (from Tables \ref{table3} and \ref{table4}). Specifically, estimated DEA scores excluding outliers tend to under-estimate efficiency levels in the case of cognitive efficiency models (in public schools by -1.68\% and in private schools by -1.05\%) as well as for the private school non-cognitive model (by -0.88\%), only over-estimating school's efficiency for the non-cognitive outcome public school model (by +2.33\%).

\begin{table}[htbp]
\begin{threeparttable}
\caption{Comparison of DEA school efficiency estimates and those excluding supper efficient schools}
\begin{tabular}{llllll}
\hline
 & \multicolumn{1}{c}{Public schools} &  &  & \multicolumn{1}{c}{Private schools} \\
    \cmidrule{2-3} \cmidrule{5-6}
 & Efficiency difference (\%) & $\Delta$ N &  & Efficiency difference (\%) & $\Delta$ N \\
 \hline
 & (1) & (2) &  & (3) & (4) \\
\textit{Panel A – Cognitive outcomes} &  &  &  &  \\
Whole sample & -1.68 & -261 &  & -1.05 &-282 \\
Argentina & -2.28 & -48 &  & 0.45 & -60 \\
Brazil & -2.12 & -60 &  & -4.01 & -36 \\
Chile & -1.81 & -15 &  & -0.52 & -62 \\
Colombia & -0.89 & -24 &  & -0.45 & -24 \\
Dominican Rep. & 1.05 & 0 &  & 3.04 & -1 \\
Mexico & -0.68 & -23 &  & 0.42 & -20 \\
Panama & 0.02 & -3 &  & -1.29 & -12 \\
Peru & -1.02 & -36 &  & 1.42 & -46 \\
Uruguay & -3.73 & -52 &  & -2.06 & -21 \\
 &  &  &  &  \\
\textit{Panel B – Non-cognitive outcome} &  &  &  &  \\
Whole sample & 2.33 & -149 &  & -0.88 & -343 \\
Argentina & 2.14 & -16 &  & -0.29 & -74 \\
Brazil & 3.37 & -13 &  & -3.09 & -34 \\
Chile & -0.03 & -25 &  & 3.01 & -84 \\
Colombia & 2.91 & -11 &  & -2.41-33 \\
Dominican Rep. & 2.28 & -19 &  & -9.53 & -23 \\
Mexico & 2.39 & -22 &  & -4.56 & -21 \\
Panama & 2.00 & -8 &  & -1.17 & -17 \\
Peru & 2.60 & -16 &  & 2.40 & -38 \\
Uruguay & 2.52 & -19 &  & 2.16 & -19 \\
\hline
\end{tabular}
\label{tableC1}
\begin{tablenotes}[para,flushleft]
\medskip
\footnotesize{
(1) Proportional changes on efficiency score are obtained as the proportion of the gap of the estimate efficiency scores of Tables \ref{table3} and \ref{table4} with those excluding supper efficient schools.
}
\end{tablenotes}
\end{threeparttable}
\end{table}


\renewcommand{\thesubsection}{\Alph{subsection}}
\setcounter{figure}{0}
\renewcommand{\thefigure}{D\arabic{figure}}
\setcounter{table}{0}
\renewcommand{\thetable}{D\arabic{table}}

\begin{table}[htbp]
\caption{Efficiency determinants definitions}
\begin{tabular}{ll}
\hline
Variable name & Definition  \\
\hline
\textit{Panel A – student variables} \\
StudBKGD\_Gender  & Gender (male) \\
StudBKGD\_Homework  & Homework in all subjects-time spent \\
StudBKGD\_Books  & Number of books at home \\
StudBKGD\_OtherLang  & Speak other language  \\
StudBKGD\_Immigrant  &  Comes from migrant family \\
StudBKGD\_WorkUnpaid  &  Work unpaid days \\
StudBKGD\_WorkPaid  & Work paid days \\
StudBKGD\_RepePrim  & Repeated primary \\
StudBKGD\_RepeLowsec  & Repeated lower secondary \\
StudBKGD\_MissClass & Skipped classes \\
StudBKGD\_BelongSch  & Sense of belonging to school \\
StudBKGD\_SchClimaBad  & Bad school climate \\
StudBKGD\_TchSuppMath  & Teachers support in class \\
StudBKGD\_ICTHomeResources  & ICT home resources \\
StudBKGD\_ICTResourcesUse  & ICT resources use \\
StudBKGDCovid\_OwnPChHomework   & Covid: device used during school closure \\
StudBKGDCovid\_ProbHomeworkWeb  & Covid: problem schoolwork – access and connectivity \\
StudBKGDCovid\_ProbHomeworkHelp  & Covid: problem schoolwork – understanding and no help \\
 & \\
\textit{Panel B – school variables} \\
SchBKGD\_Urban  & Urban  \\
SchBKGD\_Competition  & Competition \\
SchBKGD\_AdmissionPerfo  & Admission - academic record \\
SchBKGD\_AdmissionResi  & Admission – area of residence \\
SchBKGD\_AdmissionReli  & Admission - religious philosophy \\
SchBKGD\_TchPDAtt  & Teachers attended PD programme in last 3 months \\
SchBKGD\_TchCertified  & Certified teachers  \\
SchBKGD\_TchMonitor  & Methods to monitor the practice of teachers \\
SchBKGD\_PolSocEmSkill  & Social and emotional skills policy \\
SchBKGD\_QualPol  & Improvements and quality policies \\
SchBKGD\_Autonomy  & Autonomy \\
SchBKGD\_HinderInstruMate  & Instruction hindered by poor educational material \\
SchBKGD\_HinderInstruInfra  & Instruction hindered by poor infrastructure \\
SchBKGD\_HinderInstruICT  & Instruction hindered by poor ICT \\
SchBKGD\_NoLangTest  & Students with a second language \\
SchBKGD\_Disadvantaged  & Disadvantaged students \\
 & \\
\textit{Panel C – COVID school variables} \\
SchBKGDCovid\_PolResourStudLearn  & Covid: resources to support students learning \\
SchBKGDCovid\_PolTchCommStudLearn  & Covid: teacher communication w/ students and parents \\
SchBKGDCovid\_PolSuppTchRemoteL  & Covid: resources to support teachers for remote learning \\
SchBKGDCovid\_PropStudRemoteL  & Covid: students attending distance learning \\
SchBKGDCovid\_TchSkillsRemoteL & Covid: teachers have skills for instruction \\
SchBKGDCovid\_TchTimeLessonDD  & Covid: teachers have time to integrate digital devices in lessons  \\
SchBKGDCovid\_TchOnlinePlatform  & Covid: teachers have learning support platform \\
SchBKGDCovid\_BarrierRemoteLDD  & Covid: remote instruction hindered by digital devices access \\
SchBKGDCovid\_BarrierRemoteLWeb  & Covid: remote instruction hindered by internet access \\
SchBKGDCovid\_BarrierRemoteLSysEM  & Covid: remote instruction hindered by system and material \\
SchBKGDCovid\_Closeddays  & Covid: number of days school closed \\
\hline
\end{tabular}
\label{tableD1}
\end{table}

\begin{figure}[ht!]
\centering
\subfloat[\footnotesize{Cognitive outcomes-public schools}]{\includegraphics[width=0.49\textwidth]{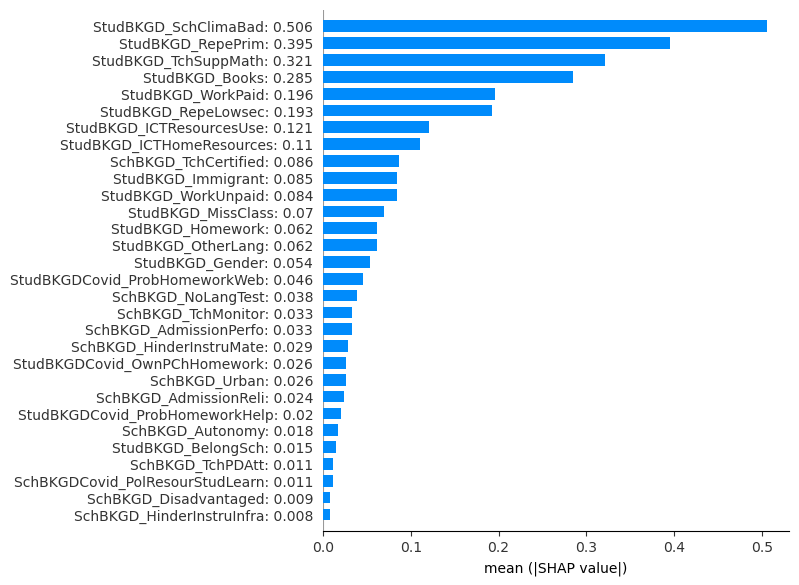}\label{figureD1a}}
\subfloat[\footnotesize{Cognitive outcomes-private schools}]{\includegraphics[width=0.49\textwidth]{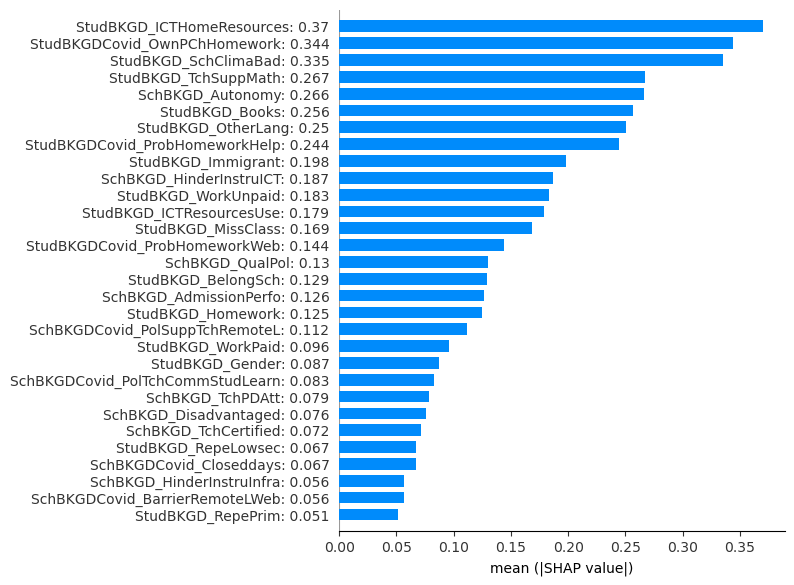}\label{figureD1b}}
\\
\subfloat[\footnotesize{Non cognitive outcomes-public schools}]{\includegraphics[width=0.49\textwidth]{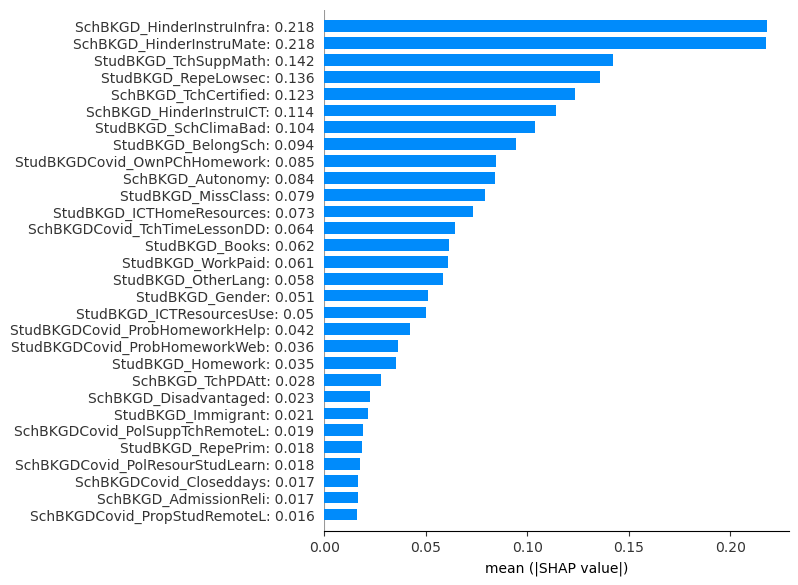}\label{figureD1c}}
\subfloat[\footnotesize{Non cognitive outcomes-private schools}]{\includegraphics[width=0.49\textwidth]{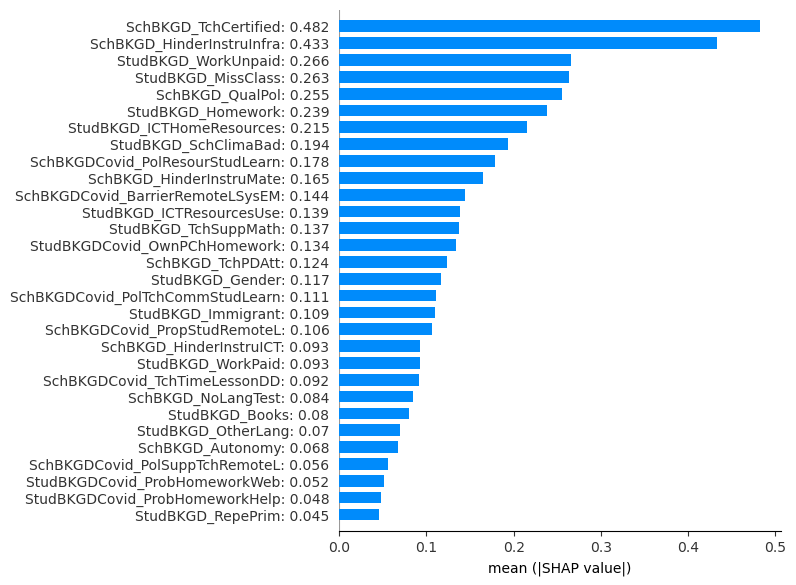}\label{figureD1d}}
\caption{SHAP values estimates of top determinants of school efficiency}
\label{figureD1}
\medskip
\begin{minipage}{0.99\textwidth}
\footnotesize{
Notes: (1). Importance plots show ranked average absolute SHAP values per covariate (top 30 covariates). (2) Details about covariates' definition can be found in Table \ref{tableD1}.
}
\end{minipage}
\end{figure}

\end{document}